\documentclass[a4paper,fleqn]{cas-sc}
\usepackage[numbers]{natbib}
\usepackage{algorithm}
\usepackage{algorithmic}
\usepackage{amsmath,amssymb}
\usepackage{amsthm}
\usepackage{booktabs}
\usepackage{graphicx}
\usepackage{multirow}
\usepackage{xcolor}
\usepackage{url}
\usepackage{enumitem}
\usepackage{xspace}

\usepackage{graphicx}
\usepackage{caption}
\usepackage{subcaption}

\usepackage[section]{placeins}

\hypersetup{
    colorlinks=true,
    citecolor=blue,
    linkcolor=blue,
    urlcolor=blue
}

\newcommand{\SystemName}{ElastiCo\xspace}

\newcommand{\mypara}[1]{\smallskip\noindent\textbf{#1}}

\begin{document}

\let\SavedPagebreak\pagebreak
\renewcommand{\pagebreak}[1][]{}

\shorttitle{ElastiCo: Elastic Configuration and Interference-Aware Orchestration}

\shortauthors{J.~Wang et~al.}

\title[mode=title]{ElastiCo: Elastic Configuration and Interference-Aware Orchestration for GPU Clusters}

\author[1]{Jinghao Wang}
\ead{wang_jinghao@buaa.edu.cn}
\author[2]{Yihang Zhou}
\ead{zhou_yihang@buaa.edu.cn}
\author[3]{Xiaoyang Sun}
\cormark[1]
\ead{X.Sun4@leeds.ac.uk}
\author[1]{Chunming Hu}
\ead{hucm@buaa.edu.cn}
\author[1]{Tianyu Wo}
\ead{woty@buaa.edu.cn}
\author[1]{Xu Wang}
\ead{xuwang@buaa.edu.cn}
\author[4]{Albert Y. Zomaya}
\ead{albert.zomaya@sydney.edu.au}
\author[1]{Renyu Yang}
\ead{renyuyang@buaa.edu.cn}
\cortext[cor1]{Corresponding author}

\affiliation[1]{organization={School of Software, Beihang University},
    city={Beijing},
    postcode={100083},
    country={China}}
\affiliation[2]{organization={School of Computer Science and Engineering, Beihang University},
    city={Beijing},
    postcode={100083},
    country={China}}
\affiliation[3]{organization={School of Computer Science, University of Leeds},
    city={Leeds},
    postcode={LS2 9JT},
    country={United Kingdom}}
\affiliation[4]{organization={School of Computer Science, University of Sydney},
    city={Sydney},
    country={Australia}}

\begin{abstract}
Modern GPU clusters must simultaneously serve deep learning training and offline large language model inference workloads, yet existing schedulers treat these as isolated resource consumers with rigid, static allocations. This leaves substantial GPU capacity underutilized: training jobs reserve entire devices despite periodic idle phases, while offline inference tasks over-provision GPUs despite bursty demand patterns. 
We present \SystemName, an elastic co-location framework that enables training and inference workloads to safely share GPUs through three integrated mechanisms. First, \emph{Resource Shape Transformation} exposes each job as a family of feasible resource-performance configurations. Second, \emph{Elastic Shadow Pricing} decomposes the resulting multi-resource allocation problem into per-job configuration selection subproblems via dynamic per-resource shadow prices. Third, \emph{Interference-Aware Co-location} uses a predictor trained on hardware-counter and task-level features to estimate pairwise performance degradation under GPU sharing.  
Implemented as native Kubernetes middleware requiring no user-code modifications, \SystemName is evaluated on a 64-GPU testbed and through large-scale trace-driven simulations (up to 512 GPUs), reducing the average JCT by up to 2.94$\times$, increasing the cluster throughput by 2.02$\times$, and increasing the GPU utilization from approximately 25\% to 46\%.
\end{abstract}


\maketitle

\let\pagebreak\SavedPagebreak

\section{Introduction}
\label{sec:introduction}

GPU clusters are the single largest upfront investment in contemporary AI infrastructure, yet real-world production data repeatedly expose a challenge: even under nominal over-subscription, average GPU utilization almost never exceeds 30\%~\cite{chen2025mudi, weng2022mlaas, gao2024lowgpu}. This kind of waste is not an issue of workload, but a system \emph{architectural} design. Modern compute clusters are increasingly required to handle two primary types of workloads: deep learning (DL) training, which uses large numbers of GPUs for many hours or even days to run iterative gradient descent, and \emph{offline LLM inference}, which covers batch-style jobs such as large-scale data labeling, model evaluation, synthetic data creation, etc.~\cite{hu2024characterization, weng2022mlaas}. In contrast to latency-sensitive online serving, offline inference focuses on maximizing overall throughput rather than minimizing the latency of individual requests, but it still requires substantial GPU compute and memory. As demand for both types of workload increases, the cluster must handle them at the same time, turning efficient resource sharing into a primary infrastructure challenge.

The primary reason for this underutilization is \emph{static resource partitioning}: operators allocate distinct GPU pools for training and inference, and provision each pool according to its own peak demand. Existing studies~\cite{chen2025mudi, gao2024lowgpu} show that, in real production clusters, the average utilization of GPU SMs typically stays well below 30\%.
An increasing number of studies show that running training and inference together on the same GPUs can greatly boost utilization, because these two types of workloads have bursty, mostly non-overlapping resource demands that introduce exploitable idle periods~\cite{lv2025specinf, wang2025sirius, liu2025smore, chen2024gpucolo}.
However, fully unlocking this potential demands overcoming three tightly interrelated challenges that current systems address only in isolation.

\textit{Challenge-1: Existing schedulers assume that each job has a fixed resource requirement} and thus fail to leverage the significant \emph{configuration elasticity} that DL workloads naturally provide. For instance, a training job can exchange increased computation for reduced memory usage by using activation checkpointing.
\textit{Challenge-2: Existing cluster schedulers}, like Pollux~\cite{qiao2021pollux} and Lucid~\cite{hu2023lucid}, \textit{focus solely on optimizing GPU allocation or packing for training jobs}, and do not incorporate inference workloads or co-location interference into their models. 
\textit{Challenge-3: Existing interference predictors for pairwise co-location~\cite{liu2025smore, chen2025mudi, lv2025specinf, wang2025sirius} are designed for specific co-location patterns and fail to capture cluster-wide, multi-resource optimization}. Heuristic approaches also exhibit poor scalability as the configuration space expands. Moreover, even with carefully selected configurations and allocations, co-located workloads still compete for shared hardware resources, such as streaming multiprocessors, memory bandwidth, and on-device memory.

These three challenges are interdependent: static configuration constrains which co-location setups are even possible, allocation complexity affects whether the best setup is selected, and interference awareness guarantees that the chosen setups remain safe during execution. Addressing any challenge on its own delivers only limited benefits because the remaining unresolved challenges quickly become the dominant bottlenecks.

We present \SystemName, an elastic co-location framework that simultaneously tackles all three of these challenges.
The rationale is that configuration flexibility, resource allocation, and interference awareness are interdependent aspects of one optimization problem and therefore need to be designed jointly.
\SystemName introduces \emph{Resource Shape Transformation} (RST, \S\ref{sec:methodology:rst}) to expose each job as a family of feasible resource--performance profiles, \emph{Elastic Shadow Pricing} (ESP, \S\ref{sec:methodology:esp}) to solve the resulting multi-resource allocation via cost/price-guided constraints, and \emph{Interference-Aware Co-location} (IAC, \S\ref{sec:methodology:iac}) to predict and bound pairwise degradation. 
\emph{Phase-Aware Disaggregated Scheduling} (PDS, \S\ref{sec:methodology:pds}) orchestrates these components in a closed-loop control cycle, coordinating queue-aware capacity reservation, configuration switching, and preemptive migration. 

\SystemName is realized as Kubernetes-native middleware and operates without requiring any changes to user code.
\SystemName is evaluated on a 64-GPU A100 testbed and via large-scale trace-driven simulations (scaling up to 512 GPUs). 
The results indicate that \SystemName reduces average job completion time (JCT) by up to 2.94$\times$, boosts cluster throughput by 2.02$\times$, increases GPU utilization from about 25\% to 46\%, and decreases the number of extra GPU instances needed for concurrent offline inference by 44\%.

To conclude, this paper offers the following key contributions:

\begin{enumerate}[leftmargin=*, nosep]
    \item \textbf{Resource Shape Transformation (RST)}, a method that models each job as a family of resource-performance profiles by exploring systematic settings, such as activation checkpointing, micro-batch sizing, mixed-precision modes, and KV-cache configurations (\S\ref{sec:methodology:rst}). To our knowledge, RST is the first approach that treats intra-job configuration flexibility as a first-class scheduling dimension for job co-location.

    \item \textbf{Elastic Shadow Pricing (ESP)}, a cost-based mechanism that decomposes joint configuration selection and multi-resource assignment into per-job subproblems with interference penalties, scaling to hundreds of jobs (\S\ref{sec:methodology:esp}).

    \item \textbf{Interference-Aware Co-location (IAC)}, an interference predictor over hardware-counter and task-level features, combined with adaptive tolerance thresholds, that enables interference-aware co-location admission and preemptive migration to preserve per-job performance under aggressive GPU sharing (\S\ref{sec:methodology:iac}).

    \item \textbf{Production-Grade Implementation}, a Kubernetes-native middleware with non-intrusive profiling via a two-dimensional optimized profiler, framework-native memory budget coordination (\S\ref{sec:implementation}), and a comprehensive evaluation demonstrating up to 2.94$\times$ JCT reduction, 2.02$\times$ throughput improvement, and 44\% fewer GPU instances compared to static partitioning (\S\ref{sec:evaluation}).
\end{enumerate}

\section{Background and Motivation}
\label{sec:motivation}

This section measures the resource inefficiencies of current GPU cluster management, highlights the resulting opportunities for workload co-location, and outlines the design challenges that \SystemName is built to overcome.

\subsection{GPU Underutilization in Production}
\label{sec:motivation:underutil}

\begin{figure}[pos=t]
  \centering
  \includegraphics[width=0.9\columnwidth,height=0.2\textheight,keepaspectratio]{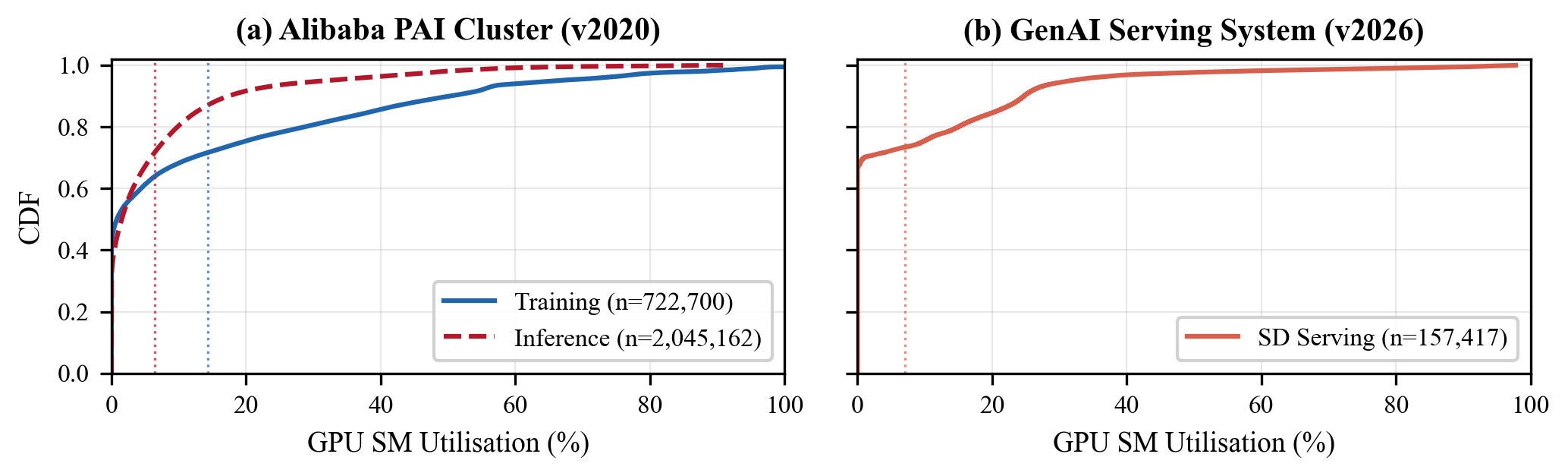}
  \caption{CDF of GPU SM utilization across two production systems.
  (a)~Alibaba PAI Cluster~\cite{weng2022mlaas}: training vs.\ inference jobs.
  (b)~GenAI Serving System~\cite{yan2025understanding}: per-pod time-series samples.}
  \label{fig:util_cdf}
\end{figure}

Two real-world production datasets have been made available: the Alibaba PAI cluster~\cite{weng2022mlaas} and a stable diffusion serving platform~\cite{yan2025understanding}. They provide two months of training and inference traces, in terms of GPU SM utilization.

Figure~\ref{fig:util_cdf} shows the cumulative distribution of GPU SM utilization. In the PAI cluster, training jobs (which request a fixed number of GPUs) reach an average utilization of only 14.3\%, and nearly 90\% of them operate below 50\%. Inference jobs (which obtain short-term GPUs on demand) perform even worse, with a mean SM utilization of just 6.4\%. Overall, the cluster-wide average is 10.5\%, indicating that most allocated GPU cycles remain unused. The GenAI serving system follows a comparable trend: its GPU pods achieve an average utilization of 7.0\%, with 97.7\% of time-series measurements under 50\% and a median that is effectively zero, pointing to long idle periods between request bursts. These results align with recent industry-wide reports that production GPU clusters typically exhibit average utilizations below 30\%~\cite{chen2025mudi, gao2024lowgpu}.

This waste stems from several sources, such as naturally bursty workloads, peak-based resource provisioning, and - most importantly - \emph{static pool partitioning}, which partitions training and inference into distinct GPU pools. Because low usage in one pool cannot offset surges in the other, a large portion of reserved hardware remains unused, even as other jobs are left without enough compute.

\subsection{Divergent Demand Patterns}
\label{sec:motivation:burstiness}

\begin{figure}[pos=t]
  \centering
  \includegraphics[width=0.9\columnwidth,height=0.2\textheight,keepaspectratio]{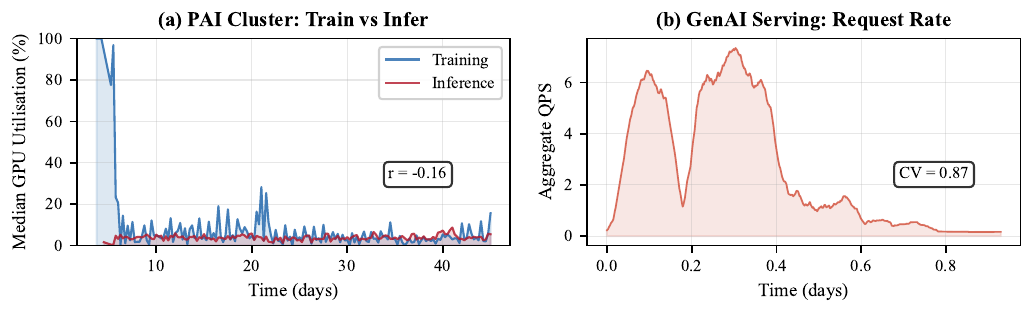}
  \caption{Temporal demand characteristics.
  (a) Median GPU utilization in the PAI cluster, aggregated over 6-hour intervals;
  (b) Total QPS of the GenAI serving infrastructure.}
  \label{fig:temporal}
\end{figure}

A co-location strategy can exploit this underutilization only when the idle periods are \emph{accessible}. As demonstrated, training and inference workloads have fundamentally different demand profiles: both are bursty, but in distinct ways, creating exactly this kind of exploitable slack.

Figure~\ref{fig:temporal}(a) shows the median GPU utilization of training and inference jobs in the PAI cluster over 6-hour intervals across a representative 45-day period.
Training experiences an early spike (around 90\%) followed by a long tail of low utilization (5--15\%), while inference stays consistently below 10\%. As a result, neither class maintains high utilization for long, and GPUs remain underused for most of the trace.
The Pearson correlation between the two time series is weakly negative ($r = -0.16$, $p < 0.05$), indicating only a mild inverse association.
Importantly, co-location does not rely on strong anti-correlation; it is enough that the two workloads \emph{infrequently peak together}, so that one pool typically has slack when the other is heavily loaded.

The GenAI serving trace supports this conclusion. Figure~\ref{fig:temporal}(b) shows the aggregate request rate (QPS) for the Stable Diffusion system, which varies sharply, with a coefficient of variation of 0.87. The peak-to-trough ratio is nearly 85$\times$ (comparing the 95th to the 5th percentile of non-zero QPS), and QPS frequently falls close to zero for prolonged periods between bursts. Each such idle stretch is a period during which GPUs reserved for inference are completely unused - this is exactly the temporal slack.

\subsection{Opportunity for Reconfigurability}
\label{sec:motivation:elasticity}

Production traces show that workloads have substantial, but largely untapped, configuration flexibility. Bursty demand dictates \emph{when} co-location is possible, while configuration elasticity determines \emph{how much} co-location can occur.

\begin{figure}[pos=t]
  \centering
  \includegraphics[width=0.9\columnwidth,height=0.2\textheight,keepaspectratio]{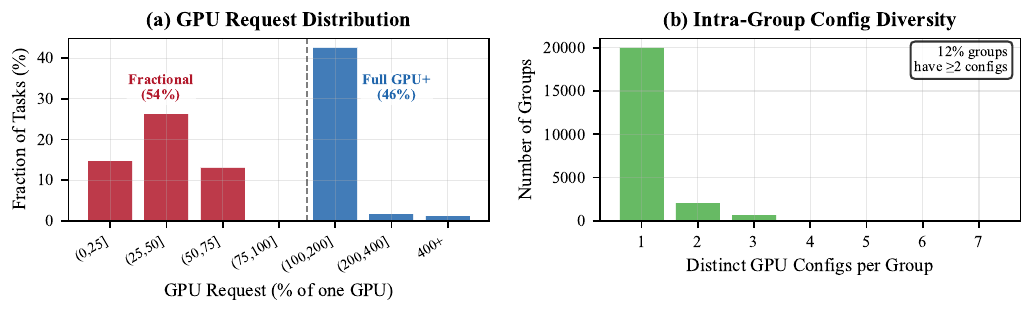}
  \caption{Job states in the PAI cluster.
  (a)~Distribution of GPU requests;
  (b)~Intra-group configuration diversity.}
  \label{fig:elasticity}
\end{figure}

\begin{figure}[pos=htbp]
  \centering
  \includegraphics[width=0.9\columnwidth,height=0.2\textheight,keepaspectratio]{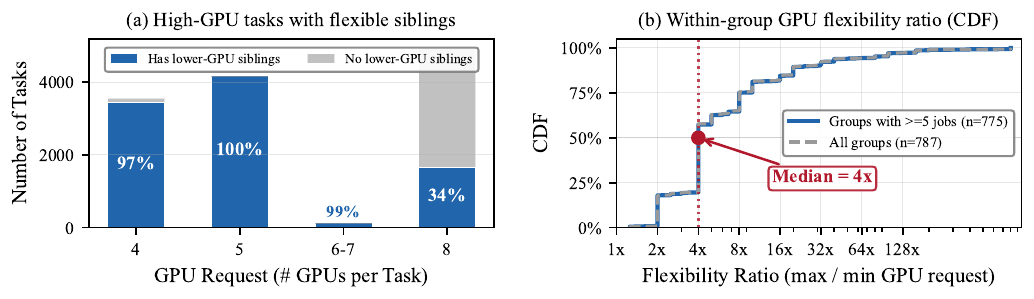}
  \caption{Job configurations in the PAI cluster.
  (a)~GPU request vs usage.
  (b)~CDF of within-group GPU flexibility ratio.}
  \label{fig:flex_evidence}
\end{figure}

Figure~\ref{fig:elasticity}(a) presents the distribution of GPU requests in the PAI cluster. Over 54\% of tasks request \emph{fractional} GPUs, indicating that many workloads can run without exclusive access to a full GPU. Figure~\ref{fig:elasticity}(b) examines workload \emph{groups} - sets of jobs with identical scripts, parameters, and data sources - and shows that among recurring workload families (groups with $\geq$5 jobs), 12\% use two or more distinct GPU configurations, with some spanning up to 7, demonstrating recurring configuration heterogeneity in production. While the group-level view highlights diversity across workload families, a complementary \emph{user-level} analysis, Figure~\ref{fig:flex_evidence}, estimates how many individual high-GPU tasks could instead run at smaller scales.

This diversity offers indirect evidence of configuration flexibility: when jobs in the same group succeed with varying GPU allocations, the workload likely supports multiple valid configurations. To characterize this, we focus on the 12,818 tasks that demand $\geq$4 GPUs and observe that 73\% were submitted by users whose overall portfolios also contain jobs completed with fewer than 4 GPUs, shown in Figure~\ref{fig:flex_evidence}(a). The median flexibility ratio per-user (max/min GPU request) is 4 times, shown in Figure~\ref{fig:flex_evidence}(b), indicating that a typical recurring workload can run over roughly a $4\times$ span of GPU counts. We further substantiate this through controlled profiling of 12 representative workloads in \S\ref{sec:evaluation:setup}, showing that each supports 12-198 feasible configurations covering different batch sizes, parallelism schemes, and memory-optimization settings.

\subsection{The Need for a Unified Orchestration Layer}

Modern GPU clusters operate on two decoupled layers - application and infrastructure. At the application level, deep learning frameworks expose configuration knobs, such as batch sizes, checkpointing policies, and KV-cache management, that determine a workload's resource footprint and performance. At the infrastructure level, cluster schedulers allocate GPUs and other resources based on simplified, typically fixed resource requests. 
This decoupling introduces three systematic challenges that limit overall cluster utilization.

\textbf{C.1 Configuration-Capacity Mismatch.}
Most production schedulers assume a job’s resource request as static throughout its entire run. However, as the trace analysis above shows, many workloads can run correctly under several valid configurations that balance performance against hardware usage. When the scheduler overlooks this flexibility, jobs can sit in the queue even though sufficient resources are actually available.
For instance, a distributed training job that requests eight GPUs for data parallelism may sit in the queue until all eight are available, even though the same job could run on four GPUs using tensor-pipeline parallelism with activation checkpointing, matching the $4\times$ median flexibility ratio seen in the trace. Since conventional schedulers are unaware of such configuration choices, they miss these opportunities, causing longer wait times and poorer hardware utilization.

\textbf{C.2 Combinatorial Allocation Complexity.}
Even when configuration choices are available, the allocation problem remains combinatorially difficult. The scheduler must simultaneously pick a configuration for every active job and distribute heterogeneous resources across potentially hundreds of concurrent workloads while maintaining both throughput and fairness.
Existing elastic training schedulers~\cite{qiao2021pollux, hu2023lucid} optimize GPU allocation for training alone without modeling inference workloads or co-location constraints, while pairwise co-location approaches do not address cluster-wide multi-resource optimization.

\textbf{C.3 Configuration-Blind Co-location.}
GPU sharing mechanisms like NVIDIA MPS enable multiple workloads to run simultaneously on one device, but current schedulers usually determine co-location feasibility using only static capacity metrics, most often GPU memory usage. In reality, however, interference between co-located workloads strongly depends on their execution configurations.
Even if two jobs fit together in memory, they can still interfere heavily at runtime by contending for shared hardware. For example, activation recomputation can greatly increase GPU compute demand, and some tensor-parallel tasks put intense pressure on SMs and memory bandwidth. Co-locating such jobs without considering this interference can severely degrade performance or even cause failures. Thus, effective co-location must account for configuration-specific hardware usage patterns, not just memory capacity.

These three challenges reveal a core shortcoming of current cluster managers: their schedulers are unaware of the \emph{configuration elasticity} of modern deep learning workloads. They cannot adapt job configurations to match available resources or anticipate interference among co-located workloads at runtime.

\section{System Design}
\label{sec:methodology}

\begin{figure}[pos=t]
\centering
\includegraphics[width=0.95\columnwidth]{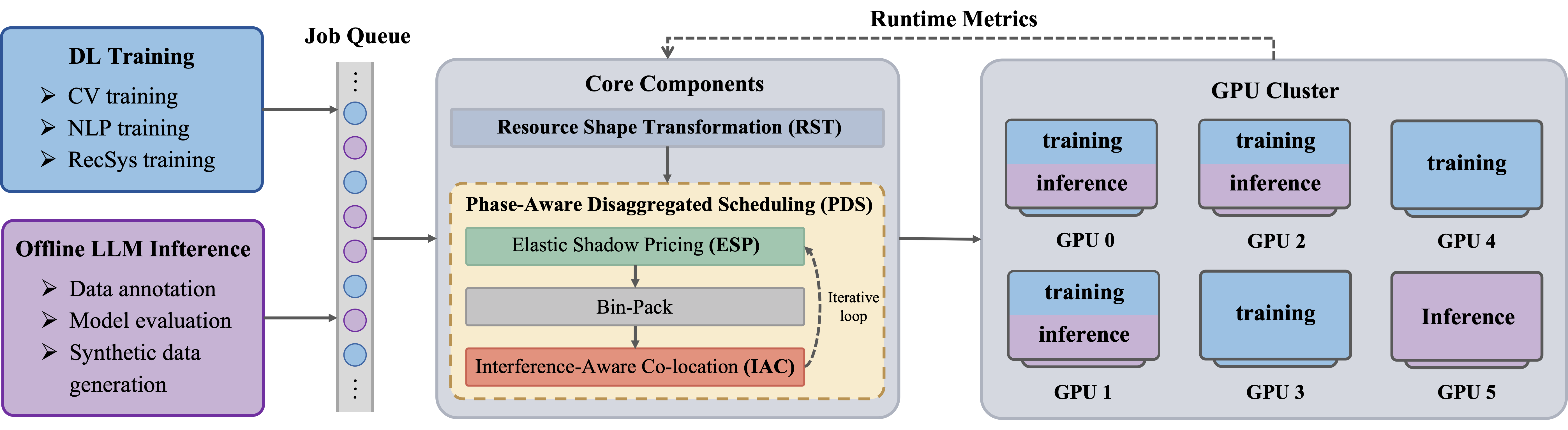}
\caption{The architecture of \SystemName.}
\label{fig:architecture}
\end{figure}

\SystemName treats \emph{configuration flexibility, resource allocation, and interference awareness} as interdependent aspects of a single optimization problem that must be designed together. Addressing any one alone turns the remaining two into hard constraints that limit utilization improvements.
This leads to a four-module architecture (Figure~\ref{fig:architecture}), where each module handles one dimension and provides clear interfaces to the others:
\begin{itemize}[leftmargin=*, nosep]
  \item \textit{Resource Shape Transformation (RST, \S\ref{sec:methodology:rst})} exposes each job as a \emph{family} of feasible resource--performance profiles, transforming rigid resource requests into a malleable configuration space.
  \item \textit{Elastic Shadow Pricing (ESP, \S\ref{sec:methodology:esp})} decomposes the combinatorial joint configuration-selection and multi-resource allocation problem into tractable per-job subproblems via a cost-efficient packing algorithm.
  \item \textit{Interference-Aware Co-location (IAC, \S\ref{sec:methodology:iac})} predicts pairwise performance degradation under GPU sharing and feeds interference costs back into ESP as penalties, internalizing co-location externalities.
  \item \textit{Phase-Aware Disaggregated Scheduling (PDS, \S\ref{sec:methodology:pds})} orchestrates the above modules in a closed-loop control cycle, coordinating runtime adaptation including queue-aware capacity reservation.
\end{itemize}
\noindent The remainder of this section formalizes each module. Table~\ref{tab:notation} summarizes the key notation used throughout.

\begin{table}[pos=htbp]
\centering\footnotesize
\caption{Notation used in the system design.}
\label{tab:notation}
\begin{tabular}{@{}ll@{}}
\toprule
\textbf{Symbol} & \textbf{Description} \\
\midrule
$\mathcal{J}$, $\mathcal{Q}$ & Active job set, pending queue \\
$\mathcal{C}_j$ & Configuration space of job $j$ \\
$R_j(c)$, $T_j(c)$, $L_j$ & Resource vector, throughput, minimum performance target \\
$\boldsymbol{\lambda}^t$ & Shadow-price vector at epoch $t$ \\
$\hat{S}_{i|j}$ & Predicted slowdown of job $i$ when co-located with $j$ \\
$\tau_j(t)$ & Adaptive tolerance threshold for job $j$ \\
$\Gamma_j^{(m)}(c)$ & Interference penalty at ESP--IAC round $m$ \\
$\mathbf{C}$, $\mathbf{C}_{\text{eff}}$ & Physical and effective cluster capacity \\
\bottomrule
\end{tabular}
\end{table}

\subsection{Resource Shape Transformation (RST)}
\label{sec:methodology:rst}
\SystemName defines the formalization of each job's configuration space. For each job $j \in \mathcal{J}$, we define a configuration set $\mathcal{C}_j = \{c_j^1, \ldots, c_j^{k}\}$. A configuration $c_j^i \in \mathcal{C}_j$ is specified by a tuple of system-level knobs: $c_j^i = (s_1, s_2, \ldots, s_n)$, for example, the micro-batch size, activation checkpointing policy, and mixed-precision mode for a training job. The RST function $RST(c_j^i)$ translates each configuration into a resource--performance profile $p_j^i = (R_j(c), T_j(c), L_j)$, where $R_j(c)$ is a multi-dimensional resource demand vector (GPU SMs, GPU memory, CPU cores, and host RAM), $T_j(c)$ is the estimated isolated performance (e.g., throughput per iteration or tokens generated per second), and $L_j$ is the minimum performance target for job $j$ (e.g., a throughput floor specified at submission time).

\subsubsection{Configuration Space}
\SystemName divides incoming workloads into two types: \textit{training jobs} and \textit{offline inference jobs}. Because they differ in execution behavior and performance bottlenecks, their resource-performance tuning knobs are managed independently.

\mypara{Training jobs}: Training jobs iterate over three computational phases (i.e., forward and backward propagation, and parameter update) for thousands or millions of iterations until convergence. 
RST focuses on \textit{system-level configuration knobs} that reshape the job's resource footprint without altering model semantics. The settings are as follows.

\textit{(i) Activation Recomputation}~\footnote{Activation recomputation (also known as activation checkpointing~\cite{chen2016training}) trades additional computation for reduced memory usage.}. Activation recomputation creates a tunable memory-compute trade-off: disabling recomputation maximizes throughput at the cost of higher GPU memory consumption, while enabling it significantly reduces the peak memory footprint by recomputing activations during the backward pass.
    
\textit{(ii) Micro-Batch Size}. The micro-batch size is the number of training samples processed per GPU. It directly controls the trade-off between memory consumption and training throughput. Larger micro-batches increase arithmetic intensity and per-iteration throughput but demand proportionally more memory for activations and intermediate tensors; smaller micro-batches reduce peak memory at the cost of lower computational efficiency. 

\textit{(iii) Automatic Mixed Precision (AMP)}. Mixed-precision training~\cite{micikevicius2018mixed} executes the forward and backward passes in half-precision arithmetic (FP16 or BF16) while maintaining a master copy of the weights in FP32 for numerically stable gradient accumulation. Enabling AMP roughly halves the memory footprint of activations and intermediate tensors and increases arithmetic throughput on hardware equipped with dedicated half-precision units, with negligible impact on final model quality. 

\mypara{Offline inference jobs}:
Offline inference jobs, such as large-scale dataset annotation or model evaluation, prioritize aggregate throughput rather than per-request latency. Their performance is primarily determined by batching efficiency and memory usage associated with attention key-value (KV) caches.
RST focuses on the following configuration knobs, whose interactions collectively determine the memory-throughput trade-off.

\textit{(i) Prefix Caching}~\cite{zheng2024sglang}. LLM inference maintains a KV cache storing attention states for previously processed tokens. It exploits the observation that many requests share identical prompt prefixes (e.g., system prompts or few-shot examples) by caching their KV states, thereby avoiding redundant computation. When prefix caching is disabled, each request recomputes its full prompt context; when enabled, the inference engine retains KV states for reusable prefixes across requests, improving computation reuse at the cost of higher baseline GPU memory occupancy. 
    
\textit{(ii) Continuous Batching}. Modern LLM inference engines (e.g., vLLM~\cite{kwon2023vllm}) support continuous batching, where requests dynamically join and leave an execution batch during decoding. Two parameters herein: the \textit{maximum concurrent batch size} $B_{\max}$ and the \textit{maximum sequence length} $L_{\max}$. Increasing $B_{\max}$ improves GPU utilization, while increasing $L_{\max}$ accommodates longer outputs; however, KV cache memory grows approximately as $B_{\max} \times L_{\max}$. The feasible ranges of both parameters depend on the model size and the available GPU memory budget.

\textit{(iii) GPU Memory Utilization Cap}.
Modern inference engines expose a \textit{memory utilization cap} parameter (e.g., \texttt{gpu\_memory\_utilization}) that limits the fraction of physical GPU memory the engine may occupy. A higher cap allocates more memory to KV caches and enables deeper continuous batching, thereby increasing throughput; a lower cap leaves memory headroom for co-located jobs, facilitating GPU sharing at the cost of reduced per-job throughput.

\subsubsection{Resource-Performance Profiling}
For a given job $j$ and configuration $c_j^i \in \mathcal{C}_j$, RST
constructs a resource-performance profile
$p_j^i = \bigl(R_j(c),\; T_j(c),\; L_j\bigr)$. 

\textit{(i) Resource Utilization Profiling, $R_j(c)$.} The resource vector $R_j(c)$ characterizes the hardware resources required by configuration $c_j^i$. RST measures GPU SM utilization, GPU memory footprint, host CPU usage, and memory consumption via runtime interfaces such as NVIDIA DCGM.

\textit{(ii) Performance Profiling, $T_j(c)$.} The performance metric $T_j(c)$ quantifies the execution efficiency of configuration $c_j^i$: training throughput (samples or iterations per second) for training workloads, and token-generation throughput or request completion rate for offline inference. RST additionally records auxiliary indicators such as iteration latency and GPU utilization, which are later used to estimate performance degradation under co-location.

Since deep learning training jobs typically run for hours or days, profiling the entire execution is impractical. Instead, \SystemName collects runtime statistics for a small number of iterations after the warm-up phase, during which framework initialization, memory allocation, and data pipeline setup have stabilized. Training jobs exhibit highly stable per-iteration behavior after warm-up; therefore, these sampled iterations accurately represent steady-state performance.

\textit{(iii) Execution Constraints, $L_j$.} The performance target $L_j$ captures job-specific requirements that must be satisfied during scheduling, for example, minimum throughput targets for both training and offline inference workloads. These constraints prevent the scheduler from selecting configurations that violate application-level performance objectives.

When a profile is unavailable for a particular configuration, for example, due to a hardware change or a previously unseen knob combination, RST falls back to nearest-neighbor interpolation from the profile store, selecting the entry whose resource demand vector is closest under an $\ell_2$ distance. This interpolation provides a provisional throughput estimate that is sufficient for initial scheduling decisions; once the job has been admitted, the online profiling procedure (\S\ref{subsec:profiling}) replaces the estimate with empirically measured values. 

\subsection{Elastic Shadow Pricing (ESP)}
\label{sec:methodology:esp}

\subsubsection{Problem Formulation}
Given the profile families produced by RST, the scheduler must select a configuration $c_j \in \mathcal{C}_j$ for every active job $j \in \mathcal{J}$ and decide which jobs to co-locate, so as to maximize aggregate weighted throughput subject to per-resource capacity and per-job performance constraints. We formalize this as the following integer program:

\begin{equation}
\max_{\{x_j^c\}} \;\; \sum_{j \in \mathcal{J}} \sum_{c \in \mathcal{C}_j} w_j \cdot T_j(c) \cdot x_j^c
\label{eq:primal_obj}
\end{equation}
\vspace{-1em}
\begin{align}
\text{s.t.}\quad & \sum_{c \in \mathcal{C}_j} x_j^c = 1, \quad \forall\, j \in \mathcal{J} \label{eq:one_config}\\
& \sum_{j \in \mathcal{J}} \sum_{c \in \mathcal{C}_j} R_j(c)[r] \cdot x_j^c \leq C_r, \quad \forall\, r \in \mathcal{R} \label{eq:capacity}\\
& T_j(c) \cdot x_j^c \geq L_j \cdot x_j^c, \quad \forall\, j,\, c \label{eq:perf_constraint}\\
& x_j^c \in \{0, 1\}, \quad \forall\, j,\, c \label{eq:integer}
\end{align}
where $x_j^c = 1$ if job $j$ is assigned configuration $c$ (and $0$ otherwise), $w_j$ is a per-job weight (e.g., inverse of remaining iterations for fairness), $T_j(c)$ and $R_j(c)$ are the throughput and resource vector from RST, $C_r$ is the capacity of resource $r$, and $L_j$ is the minimum throughput target for job $j$. Equation~\eqref{eq:one_config} ensures each job receives exactly one configuration; Equation~\eqref{eq:capacity} enforces cluster capacity; and Equation~\eqref{eq:perf_constraint} enforces per-job performance.

This problem is NP-hard in general.
The number of feasible solutions grows as $\prod_{j} |\mathcal{C}_j|$, which is intractable for even moderate cluster sizes. Moreover, the formulation above does not yet account for co-location interference.

\subsubsection{Lagrangian Decomposition}
Rather than solving the integer program directly, \SystemName relaxes the Equation~\eqref{eq:capacity} via Lagrange multipliers $\boldsymbol{\lambda} = (\lambda_r)_{r \in \mathcal{R}}$, yielding the Lagrangian:

\begin{equation}
\mathcal{L}(\{x_j^c\}, \boldsymbol{\lambda}) = \sum_{j} \sum_{c} \bigl[w_j T_j(c) - \textstyle\sum_{r} \lambda_r R_j(c)[r]\bigr] x_j^c + \sum_{r} \lambda_r C_r
\label{eq:lagrangian}
\end{equation}

For a fixed $\boldsymbol{\lambda}$, the Lagrangian decomposes across jobs: each job $j$ independently selects $c_j^* = \arg\max_{c} \bigl[w_j T_j(c) - \sum_r \lambda_r R_j(c)[r]\bigr]$. This per-job subproblem is trivially solvable by enumerating $|\mathcal{C}_j|$ profiles (12--198 in our evaluation; see Table~\ref{tab:workloads}). The dual function $g(\boldsymbol{\lambda}) = \max_{\{x_j^c\}} \mathcal{L}(\{x_j^c\}, \boldsymbol{\lambda})$ is then minimized over $\boldsymbol{\lambda} \geq 0$ via projected subgradient ascent, which is the economic interpretation of the t\^{a}tonnement process described below.

\subsubsection{Scoring and Price Update}
The Lagrangian decomposition motivates a practical scoring mechanism. \SystemName maintains a shadow-price vector $\lambda^t$ for each resource dimension $r\in\mathcal{R}$. The price $\lambda_r$ signals the scarcity of resource $r$ across the cluster. In each scheduling epoch, every job $j\in\mathcal{J}$ independently selects its configuration $c_j^*$ by minimizing a scoring function that corresponds to the negated per-job Lagrangian subproblem, augmented with interference and switching penalties:

\begin{equation}
    \centering
    \mathrm{Score}_{j}(c;\lambda^t) = \mathrm{PerformanceCost}_j(c) + \sum_{r \in \mathcal{R}} \lambda_r^t \cdot R_j(c)[r]+ \Gamma_j^{(m)}(c),
    \label{eq:esp_score}
\end{equation}

Here, $\mathrm{PerformanceCost}_j(c) = 1 - T_j(c)/T_j(c_j^{\max})$ measures the relative throughput loss of configuration $c$ compared with the fastest configuration $c_j^{\max} = \arg\max_{c'} T_j(c')$, where $T_j(\cdot)$ is the throughput profiled by RST (\S\ref{sec:methodology:rst}). The term $\sum_{r \in \mathcal{R}} \lambda_r^t \cdot R_j(c)[r]$ is the resource rent, where each shadow price $\lambda_r^t$ reflects the current scarcity of resource $r$. The penalty term $\Gamma_j^{(m)}(c)$ comprises two components: a \textit{switching cost} $\gamma_0 \cdot \mathbb{1}[c \neq c_j^{\mathrm{cur}}]$ that discourages unnecessary reconfiguration from the job's current configuration $c_j^{\mathrm{cur}}$, and an \textit{interference penalty} $\alpha \cdot \max(0,\, \hat{S}_{j|i} - \tau_j(t))$ injected by the IAC module (\S\ref{sec:methodology:iac}) when the predicted co-location slowdown $\hat{S}_{j|i}$ exceeds the adaptive tolerance threshold $\tau_j(t)$. The switching cost is evaluated at configuration-selection time using the job's current configuration; the interference penalty is updated after each ESP-IAC interaction round (Algorithm~\ref{alg:scheduling}, lines 11--15). 

Once all jobs have selected their configurations, the scheduler updates prices based on aggregate usage $U_r$ relative to cluster capacity $C_r$: 

\begin{equation}
    \centering
    \lambda_r^{t+1} = \mathrm{max}(0, \lambda_r^t+\eta_t(U_r-C_r))
    \label{eq:price_update}
\end{equation}
Here $U_r = \sum_{j \in \mathcal{J}} R_j(c_j^*)[r]$ is the aggregate demand for resource $r$ under the current configuration selection, and $\eta_t > 0$ is the step size (learning rate), which may be fixed or follow a diminishing schedule $\eta_t = \eta_0 / \sqrt{t}$ to balance responsiveness with stability.

This projected-subgradient update drives aggregate demand toward feasibility: if GPU memory is saturated, its price $\lambda_{\mathrm{mem}}$ rises, forcing jobs to select configurations with lower memory demand (e.g., enabling activation recomputation) until aggregate demand becomes feasible. Formally, the price update (Eq.~\ref{eq:price_update}) is a projected subgradient step on the Lagrangian dual of the capacity-constrained allocation problem (Eq.~\ref{eq:lagrangian}): the subgradient $U_r - C_r$ measures constraint violation, and projection onto $\mathbb{R}_{\geq 0}$ enforces non-negative prices.

\subsection{Interference-Aware Co-location (IAC)}
\label{sec:methodology:iac}

To increase GPU utilization, \SystemName enables multiple workloads to run on the same GPU. Although such sharing boosts overall cluster use, co-located workloads can compete for shared hardware resources-streaming multiprocessors, memory bandwidth, GPU memory, and interconnects-leading to unpredictable slowdowns if not carefully controlled.

Rather than relying on bin-packing that only accounts for static capacities, \SystemName performs \textit{interference-aware co-location}: before placing two workloads on a GPU, the scheduler estimates the performance slowdown due to resource contention. If the projected degradation exceeds a tolerance threshold, it either avoids that placement or adjusts job settings. 
The IAC module predicts the slowdown factor $\hat{S}_{i|j}$ when jobs $i$ and $j$ run together.

\subsubsection{Feature Engineering}
The IAC prediction model takes as input a feature vector characterizing a candidate co-location pair and outputs the predicted performance decay. For each job pair $(i,j)$, the model predicts $\hat{S}_{i|j} = T_i^{\mathrm{solo}} / T_i^{\mathrm{colo}}$, where $T_i^{\mathrm{solo}}$ denotes the throughput of job $i$ running in isolation and $T_i^{\mathrm{colo}}$ denotes its throughput when co-located with job $j$. A value of $\hat{S}_{i|j} = 1$ indicates zero degradation; larger values indicate proportionally greater slowdown. 

\mypara{Hardware utilization features.}
Each workload configuration is profiled in isolation to capture its steady-state resource consumption. Two metrics characterize compute pressure: \textit{SM Active}, the fraction of cycles with at least one active warp, and \textit{SM Occupancy}, the ratio of resident to maximum warps per SM. Memory subsystem behavior is captured by \textit{DRAM Active}, the fraction of cycles with active DRAM requests, and \textit{Framebuffer Used}, the GPU memory occupied in megabytes. Finally, \textit{Tensor Core Active} measures the fraction of cycles with active Tensor Core operations, distinguishing workloads dominated by matrix multiply-accumulate kernels from those that primarily execute general-purpose CUDA kernels.

\mypara{Task-level features.}
All jobs share three common attributes: a binary workload type indicator (training or inference), the log-scaled model parameter count, and the isolated throughput $T_j(c)$ from RST profiling.
Training jobs contribute three additional attributes: the micro-batch size (determining per-iteration activation memory and computation volume), an activation checkpointing flag, and a mixed-precision flag.
Inference jobs contribute the input sequence length (ISL), the average output sequence length (OSL), and the prefix caching ratio $\rho$. ISL determines the computational cost of the prefill phase and the initial KV cache allocation; OSL determines decode duration and KV cache growth. 

\mypara{Pairwise interaction features.}
Beyond per-job features, the vector includes pairwise terms capturing the combined resource pressure. The mean of each hardware utilization metric across both jobs represents aggregate contention on the corresponding resource. A co-location type indicator encodes whether the pair is training--training, training--inference, or inference--inference, as these combinations exhibit distinct interference characteristics. A memory pressure ratio (sum of both jobs' framebuffer usage divided by total GPU memory capacity $C_{\mathrm{mem}}$ (e.g., 80\,GB for A100)) quantifies proximity to the memory ceiling. A compute balance ratio $\min(\mathit{sm}_i, \mathit{sm}_j) / \max(\mathit{sm}_i, \mathit{sm}_j)$ measures the symmetry of SM demand. A memory intensity difference $|(\mathit{dram}_i / \mathit{sm}_i) - (\mathit{dram}_j / \mathit{sm}_j)|$ captures the complementarity of compute-bound and memory-bound profiles.

The complete feature vector has 36 dimensions: 14 per-job features for each job (5 hardware, 3 common, 3 training-specific, 3 inference-specific) and 8 pairwise features. Since co-located jobs generally experience different degrees of slowdown, each co-location experiment yields two training samples--one per job--with the per-job feature blocks swapped so that the first 14 dimensions always correspond to the job whose slowdown is being predicted.

\subsubsection{Interference Prediction Engine}

IAC employs a three-layer fully connected neural network to map the 36-dimensional feature vector $\mathbf{x}_{i,j}$ to the predicted slowdown $\hat{S}_{i|j}$. We adopt a DNN over tree-based alternatives for two reasons. First, the prediction target $\hat{S}_{i|j}$ is a continuous ratio that varies smoothly with resource utilization; a DNN with ReLU activations provides smooth interpolation in this space, whereas tree ensembles produce piece-wise constant predictions that can under-resolve small but scheduling-relevant slowdown differences (e.g., $1.05\times$ vs.\ $1.15\times$). Second, the training set is moderately sized (constructed from sampled co-location experiments); Batch Normalization and Dropout (rate 0.2) regularize the network effectively on such data scales, while tree ensembles with comparable depth tend to overfit on correlated hardware-counter features.

The two hidden layers compute $h_\ell = \mathrm{ReLU}(\mathrm{BN}(W_\ell h_{\ell-1} + b_\ell))$ for $\ell \in \{1, 2\}$, each followed by dropout with rate~0.2. The output layer produces:
\begin{equation}
    \hat{S}_{i|j} \;=\; 1 + \mathrm{ReLU}\!\bigl(W_3 h_2 + b_3\bigr),
    \label{eq:dnn_pred}
\end{equation}
which constrains the prediction to $\hat{S}_{i|j} \geq 1.0$, consistent with the physical invariant that co-location cannot improve isolated performance. The model is trained offline by minimizing the log-scale mean squared error:
\begin{equation}
    \mathcal{L} \;=\; \frac{1}{|\mathcal{D}|}
        \sum_{(i,j) \in \mathcal{D}}
        \bigl(\log \hat{S}_{i|j} - \log S_{i|j}\bigr)^2,
    \label{eq:dnn_loss}
\end{equation}
where $S_{i|j} = T_i^{\mathrm{solo}} / T_i^{\mathrm{colo}}$ is the ground-truth slowdown measured in offline co-location experiments. Operating in log space prevents samples with large slowdowns from dominating the gradient.

\mypara{Training data generation.}
The training dataset is constructed by executing sampled workload pairs concurrently on a shared GPU via NVIDIA MPS. For each pair, both jobs are first profiled in isolation to obtain $T_i^{\mathrm{solo}}$ and the per-job feature vector; they are then co-located under identical configurations to measure $T_i^{\mathrm{colo}}$. The ratio $S_{i|j} = T_i^{\mathrm{solo}} / T_i^{\mathrm{colo}}$ serves as the training label. In our evaluation setting, profiling 20--30\% of all candidate pair combinations--stratified by workload type (training--training, training--inference, inference--inference) to ensure each co-location category is represented--empirically provides adequate coverage for the model to generalize to held-out pairs.

The predicted $\hat{S}_{i|j}$ feeds into the ESP scoring function (Eq.~\ref{eq:esp_score}) through the interference penalty term $\Gamma_j^{(m)}(c)$, steering both configuration selection and GPU placement away from high-interference co-locations.

\subsubsection{Adaptive Tolerance Thresholds}
\SystemName permits co-location only if the predicted slowdown for all participating jobs falls within a dynamic tolerance threshold $\tau_j(t)$. This threshold adapts to the cluster contention level: when the queue is long, \SystemName tolerates slightly more interference to increase aggregate throughput.

\begin{equation}
\tau_j(t) = \tau_j^{\text{base}} + \beta \cdot \left(\frac{U(t)}{U_{\text{target}}} - 1\right)
\label{eq:adaptive_threshold}
\end{equation}
where $\tau_j^{\text{base}}$ is the baseline tolerance, $U(t)$ is the current cluster utilization, $U_{\text{target}}$ is the utilization target, and $\beta$ controls sensitivity. The baseline $\tau_j^{\text{base}}$ is set to satisfy the job's throughput constraint: $\tau_j^{\text{base}} \leq T_j(c_j^*) / L_j$, ensuring that the admitted slowdown never causes the co-located throughput to drop below the minimum target $L_j$. The additive term $\beta \cdot (\cdot)$ is clamped so that $\tau_j(t) \leq \tau_j^{\text{base}} + \beta_{\max}$, providing a hard upper bound on tolerated degradation even under extreme cluster pressure.

\subsection{Phase-Aware Disaggregated Scheduling (PDS)}
\label{sec:methodology:pds}

RST exposes configuration flexibility, ESP determines optimal configurations under multi-resource constraints, and IAC quantifies interference risk. Individually, each module addresses one dimension of the co-location problem; without coordination, however, their decisions may conflict--e.g., ESP may select a memory-intensive configuration that IAC subsequently flags as interference-prone, triggering wasteful rollbacks. The Phase-Aware Disaggregated Scheduling (PDS) module closes this loop by integrating the three modules into a unified, iterative scheduling cycle that jointly converges on resource allocations that are simultaneously capacity-feasible, interference-safe, and performance-compliant.

Specifically, the predicted slowdown factors from IAC are incorporated directly into the configuration-selection process by penalizing profiles that introduce excessive interference. For a candidate co-location pair $(i,j)$, if the predicted slowdown $\hat{S}_{i|j}$ exceeds the tolerance threshold for either workload, the scheduler reduces the effective value of the offending configuration through the penalty term in the ESP objective (Eq.~\ref{eq:esp_score}), encouraging the t\^{a}tonnement process to select alternative configurations or placements with lower interference. 

Through this feedback loop, \SystemName balances two competing goals: maximizing GPU utilization through aggressive co-location and preserving application performance by avoiding harmful contention. Rather than rejecting co-location decisions outright, the system gradually steers allocations toward interference-safe configurations.

\textbf{Scheduling Workflow.} Algorithm~\ref{alg:scheduling} summarizes the complete scheduling loop. The scheduler operates periodically at a fixed epoch interval (default $\Delta T = 5$\,s), enabling rapid reaction to workload arrivals, job completions, and configuration changes.

At the beginning of each epoch, the scheduler computes the \emph{effective resource capacity} by reserving headroom for queued jobs:
\begin{equation}
    \mathbf{C}_{\text{eff}}(t) \;=\; \mathbf{C} \;-\; \gamma \!\sum_{j \in \mathcal{Q}(t)} R_j^{\min},
    \label{eq:effective_capacity}
\end{equation}
where $\mathbf{C}$ is the physical cluster capacity, $\mathcal{Q}(t)$ is the set of pending jobs, $R_j^{\min} = \min_{c \in \mathcal{C}_j} R_j(c)$ is the smallest resource footprint in job~$j$'s profile family (from RST), and $\gamma \in [0,1]$ controls reservation aggressiveness ($\gamma{=}0$ ignores the queue; $\gamma{=}1$ reserves capacity for every pending job at its minimum footprint). By operating on $\mathbf{C}_{\text{eff}}$ rather than $\mathbf{C}$, ESP avoids over-committing resources needed for imminent job admissions, reducing reactive migrations.

\begin{algorithm}
\caption{\SystemName\ Scheduling Loop}
\label{alg:scheduling}
\begin{algorithmic}[1]
\REQUIRE Active jobs $\mathcal{J}$, queue $\mathcal{Q}$, profile sets $\{\mathcal{P}_j\}_{j \in \mathcal{J}}$, 
         capacity $\mathbf{C}$, prices $\boldsymbol{\lambda}$, max rounds $M$
\ENSURE Configuration selection $\{c_j^*\}$, placement $\mathcal{A}$, updated prices $\boldsymbol{\lambda}$
\STATE $\mathbf{C}_{\mathrm{eff}} \gets \mathbf{C} - \gamma \sum_{j \in \mathcal{Q}} R_j^{\min}$ 
       \COMMENT{Reserve capacity for queued jobs}
\STATE $\boldsymbol{\Gamma} \gets \mathbf{0}$ \COMMENT{Initialize interference penalties}
\FOR{$m = 1, \ldots, M$}
    \STATE $\boldsymbol{\lambda}^{*} \gets \textsc{EspTatonnement}(\mathcal{J}, \{\mathcal{P}_j\}, \mathbf{C}_{\mathrm{eff}}, \boldsymbol{\lambda}, \boldsymbol{\Gamma})$
    \FOR{each job $j \in \mathcal{J}$}
        \STATE $c_j^{*} \gets \arg\min_{c \in \mathcal{C}_j} \; \textsc{Score}_j(c;\, \boldsymbol{\lambda}^{*})$
    \ENDFOR
    \STATE $\mathcal{A} \gets \textsc{BinPack}\bigl(\{(j,\, R_j(c_j^{*}))\},\; \text{GPU nodes}\bigr)$
    \STATE $\mathcal{V} \gets \emptyset$ \COMMENT{Violation set}
    \FOR{each GPU $g$ with co-located pair $(i, j) \in \mathcal{A}$}
        \STATE $\hat{S}_{i|j},\, \hat{S}_{j|i} \gets \textsc{IacPredict}(i, j)$
        \IF{$\hat{S}_{i|j} > \tau_i(t)$ \textbf{or} $\hat{S}_{j|i} > \tau_j(t)$}
            \STATE Update $\boldsymbol{\Gamma}$ for offending configurations
            \STATE $\mathcal{V} \gets \mathcal{V} \cup \{(i, j)\}$
        \ENDIF
    \ENDFOR
    \IF{$\mathcal{V} = \emptyset$}
        \STATE \textbf{break} \COMMENT{All co-location pairs safe}
    \ENDIF
\ENDFOR
\STATE Apply configuration switches, placements, and migrations
\STATE Admit jobs from $\mathcal{Q}$ if residual capacity permits
\RETURN $\{c_j^*\}_{j \in \mathcal{J}},\; \mathcal{A},\; \boldsymbol{\lambda}^{*}$
\end{algorithmic}
\end{algorithm}

The scheduling loop proceeds as follows. The ESP module first computes shadow prices through a t\^{a}tonnement process based on current resource demand and capacity constraints. 
Given the resulting price vector $\boldsymbol{\lambda}^*$, each job independently selects the configuration that maximizes its net surplus--the gap between configuration value and priced resource consumption.

A bin-packing procedure then assigns jobs to GPU devices according to their resource demands. Because bin-packing considers only static capacity constraints, the resulting placement may still introduce runtime interference. The IAC module therefore evaluates every co-located pair using the interference prediction engine.

If the predicted slowdown for any pair exceeds the adaptive threshold $\tau_j(t)$, the scheduler updates the corresponding configuration penalties and records the violation. These penalties reduce the attractiveness of problematic configurations in subsequent ESP iterations. The loop repeats until no violations remain or the maximum number of rounds $M$ is reached.

\textbf{Runtime Adaptation.}
After convergence, the scheduler applies the resulting allocation decisions, which may involve switching job configurations (e.g., changing batch size or checkpointing depth), migrating workloads across GPUs, or admitting queued jobs when capacity becomes available. Because in-place configuration adjustments--propagated through Pod specification parameters (\S\ref{subsec:memory})--avoid the overhead of cross-node state transfer, PDS prioritizes reconfiguration over migration whenever possible.

\textbf{Computational Complexity.}
The per-epoch complexity of the scheduling loop is $O\!\left(M \cdot (I \cdot N \cdot \bar{K} + N^2 \cdot F)\right)$, 
where $M$ is the maximum number of ESP--IAC interaction rounds (typically 2--3), $I$ is the number of t\^{a}tonnement iterations for price convergence (typically 3--5 with warm starting), $N$ is the number of active jobs, $\bar{K}$ is the average number of profiles per job, and $F$ is the cost of a single interference prediction. 

In practice, each interference prediction takes approximately 0.08\,ms. Even for large-scale scenarios with $N=500$ active jobs and $\bar{K}=8$ profiles per job, the end-to-end scheduling latency remains under 15\,ms--negligible relative to the 5-second epoch interval--enabling \SystemName to operate efficiently at cluster scale while continuously adapting to dynamic workload conditions.

\section{Implementation}
\label{sec:implementation}

Translating the design of \S\ref{sec:methodology} into a production-grade system raises two principal engineering challenges that are not addressed by the algorithmic formulation alone: (i)~\emph{non-intrusive observability}--collecting the hardware-counter and kernel-level telemetry required by RST and IAC; and (ii)~\emph{memory isolation under co-location}--enforcing per-job GPU memory budgets when multiple workloads share a device via NVIDIA MPS. Configuration changes selected by ESP are propagated to workloads through Pod specification parameters and, when necessary, through checkpoint-based state serialization (\S\ref{subsec:memory}), ensuring safe transitions without requiring modifications to user code.

\subsection{Non-Intrusive Interception and Profiling}
\label{subsec:profiling}
 
\SystemName measures application resource usage and builds RST profiles non-intrusively, ensuring compatibility with arbitrary PyTorch, vLLM, and custom CUDA applications. This is achieved through a \emph{Two-Dimensional Optimized Profiler} (TDOP) that fuses two complementary telemetry streams: (i)~a \emph{software dimension} that intercepts CUDA driver and runtime library invocations at the user-space level to capture per-kernel launch metadata and memory allocation events, and (ii)~a \emph{hardware dimension} that continuously samples GPU performance counters via NVIDIA DCGM to characterize steady-state compute and memory subsystem behavior. The fusion of these two streams--software-level allocation traces and hardware-counter telemetry--provides richer and more accurate profiles than either stream alone, because software events reveal \emph{what} the application requests while hardware counters reveal \emph{how} the GPU responds.
 
\mypara{Instrumentation mechanism.}
The TDOP intercepts CUDA entry points via \texttt{LD\_PRELOAD}-based dynamic library interposition, inserting lightweight wrappers around \texttt{cuLaunchKernel}, \texttt{cudaMalloc}, \texttt{cudaFree}, and collective communication primitives. The wrappers record per-kernel launch metadata (grid/block dimensions, shared memory size, stream ID) and memory allocation events without modifying or delaying the underlying calls. Aggregated metrics are written to a per-process ring buffer and flushed to the data-plane agent at configurable intervals (default: every 2 seconds).
 
\mypara{Profiling.}
For each new (job type, configuration) pair, the profiling procedure proceeds in five steps:
 
\textit{(i) Launch.} The job is started under the target configuration in an isolated profiling context (a dedicated Kubernetes pod with MPS disabled to prevent interference).

\textit{(ii) Warm-up.} The job runs for a fixed warm-up budget (default: 50--100 iterations, or approximately 30 seconds, whichever comes first) to allow framework initialization, JIT compilation, memory allocator warm-up, and data pipeline prefetching to stabilize.

\textit{(iii) Measurement.} After warm-up, the profiler collects telemetry for a measurement window of approximately 60 seconds, sampling hardware counters at 1-second resolution and recording per-iteration wall-clock time and throughput. Transient outliers (e.g., from garbage collection or checkpoint I/O) are removed using a 10\%--90\% trimmed mean.

\textit{(iv) Profile construction.} The profiler computes steady-state estimates of $T_j(c)$ (throughput) and $R_j(c)$ (multi-dimensional resource vector) and derives the 36-dimensional IAC feature vector from the hardware-counter and task-level metadata. All values are written to the profile store and tagged with the current hardware signature.

\textit{(v) Stop.} The profiling pod is terminated.
 
\mypara{Amortization and caching.}
Profile generation takes 3--5 minutes per (model architecture, configuration) pair, but this cost is amortized across all jobs of the same type via the profile store. The cache hit rate in practice is high (approximately 94\% of job arrivals match an existing profile) because the set of model architectures deployed in a cluster stabilizes quickly. When a hardware upgrade or framework version change is detected via the hardware signature, only the affected profiles are invalidated, avoiding a full re-profiling cycle.
 
For unseen configurations or hardware changes, the system falls back to nearest-neighbor interpolation as described in \S\ref{sec:methodology:rst}; empirically measured profiles take precedence once available.

\subsection{Configuration Propagation and Memory Budget Coordination}
\label{subsec:memory}

\SystemName is a scheduler-level system: it decides \emph{what} configuration
each job should run under and \emph{where} the job should be placed, but it
does not replace or bypass the memory allocators provided by deep learning
frameworks or the GPU driver.  All resource budgets are enforced by setting
the appropriate launch parameters in the Pod specification before the job
starts (or restarts after a heavy reconfiguration).  This section describes
how those parameters are chosen and propagated.

\mypara{Memory budget derivation.}
For each job~$j$ assigned to configuration~$c_j^*$ by ESP, the scheduler
computes a memory budget $M_j = R_j(c_j^*)\![\texttt{GPU\_mem}]$ from the RST
profile.  When two jobs $j_1, j_2$ are co-located on the same GPU by PDS,
the scheduler verifies the packing constraint
$M_{j_1} + M_{j_2} \le M_{\text{GPU}} - M_{\text{sys}}$, where
$M_{\text{GPU}}$ is the total device memory and $M_{\text{sys}}$ is a
reserved margin for driver and MPS overhead (default: 512\,MB).

\mypara{Framework-specific parameter mapping.}
The computed memory budget is translated into framework-native parameters and
injected into the Pod specification as environment variables or container
arguments:

\textit{(i) vLLM inference.}  The budget is mapped to \texttt{gpu\_memory\_utilization} $= M_j / M_{\text{GPU}}$, which controls the fraction of device memory that the vLLM engine pre-allocates for KV-cache blocks and model weights.

\textit{(ii) PyTorch training.}  The budget is mapped to \texttt{PYTORCH\_CUDA\_ALLOC\_CONF=\allowbreak max\_split\_size\_mb:}$M_j$ and, where supported, to \texttt{torch.cuda.set\_per\_process\_memory\_fraction}$(M_j / M_{\text{GPU}})$, which caps the CUDA caching allocator's consumption.
      
\textit{(iii) MPS resource limits.}  When NVIDIA MPS is enabled on the node (a cluster-level infrastructure decision outside \SystemName's scope), the  \texttt{CUDA\_MPS\_PINNED\_DEVICE\_MEM\_LIMIT} environment variable is set per client process to enforce a hard memory ceiling at the driver level.

\noindent
Because each framework's allocator operates within the externally imposed
ceiling, co-located jobs are isolated without requiring a custom GPU memory
manager.  This \emph{budget-from-above} design leverages
the memory-capping interfaces that frameworks already expose, translating
scheduler-computed budgets into framework-native parameters.  This preserves the non-intrusiveness guarantee (\S\ref{sec:implementation}) while achieving precise memory isolation under co-location.

\mypara{State serialization for migration.}
When a job must be migrated--e.g., because IAC identifies a harmful
co-location pair at runtime--\SystemName triggers a framework-native
checkpoint through the data-plane sidecar agent.  For PyTorch training jobs this invokes
\texttt{torch.save()} on model parameters and optimizer states; for vLLM
inference services the engine is gracefully drained and the model weights
are already resident in host memory or can be reloaded from the model
repository.  The checkpoint is written to a shared persistent volume
(NVMe-backed by default) accessible from any node, so the replacement pod can
restore without data movement across the network.  Serialization runs
asynchronously: the scheduling epoch is blocked only for the brief final
consistency flush (typically $<$\,1\,s for a 7\,B-parameter model).

\section{Evaluation}
\label{sec:evaluation}

We evaluate \SystemName through testbed experiments on a 64-GPU cluster and large-scale trace-driven simulations (up to 512 GPUs), addressing six questions:
\begin{enumerate}[nosep, label=\textbf{Q\arabic*}]
  \item How does \SystemName compare to state-of-the-art baselines in JCT, throughput, and utilization? (\S\ref{sec:evaluation:comparison})
  \item How does each component (RST, ESP, IAC, PDS) contribute to overall performance? (\S\ref{sec:evaluation:ablation})
  \item What scheduling behaviors produce the observed gains? (\S\ref{sec:evaluation:case})
  \item How accurate is the IAC interference predictor? (\S\ref{sec:evaluation:prediction})
  \item How does \SystemName scale with cluster size and load intensity? (\S\ref{sec:evaluation:scalability})
  \item What system overheads does \SystemName incur? (\S\ref{sec:evaluation:overhead})
\end{enumerate}

\subsection{Experimental Setup}
\label{sec:evaluation:setup}

\mypara{Testbed.} 8 nodes, each with 8 NVIDIA A100-40GB GPUs (64 GPUs total), interconnected via 200\,Gbps InfiniBand HDR with intra-node NVSwitch. All nodes run Ubuntu 22.04 with CUDA 12.8, PyTorch 2.8, and vLLM 0.11.0. 

\mypara{Workloads.} Table~\ref{tab:workloads} lists the 12 workloads used in our evaluation, spanning eight training tasks and four offline inference tasks. Training workloads cover diverse model architectures: image classification (ResNet-50), language understanding (BERT), generative modeling (DCGAN), 3D point cloud processing (PointNet), sequence modeling (Transformer, LSTM), recommendation (NCF-NeuMF), and reinforcement learning (PPO). Inference workloads target four LLMs of varying scale: DeepSeek-1.5b, Qwen-1.7b, Mistral-7b, and LLaMA-3-7b. Each workload is profiled under multiple configurations by varying batch size, activation checkpointing, mixed-precision training, sequence length, and prefix caching, yielding diverse feasible profiles per workload as detailed in the table. All workloads are executed on the physical testbed described above.

\mypara{Simulation.} For large-scale experiments beyond the 64-GPU testbed, we use a discrete-event simulator calibrated against testbed measurements. The simulator models GPU resource contention, MPS overhead, and interference using the IAC predictor. We validate the simulator by replaying the 24-hour testbed trace at 64 GPUs and comparing JCT and throughput: simulated results match testbed measurements to within 7\%, confirming fidelity. The simulator is used for scalability experiments at 128, 256, and 512 GPUs (\S\ref{sec:evaluation:scalability}).

\begin{table}[pos=t]
\centering
\caption{Workloads used in evaluation. Each workload is profiled under
multiple configurations by varying the listed knobs; the resulting
Cartesian product defines the set of feasible profiles per workload.}
\label{tab:workloads}
\begin{tabular}{@{}llcl@{}}
\toprule
\textbf{Workload} & \textbf{Type} & \textbf{Mem.\,(GB)} & \textbf{Configuration Knobs} \\
\midrule
ResNet-50~\cite{he2016resnet}    & Train & 12--24 & BS$\in$\{32,64, $\cdots$,512\},\; CK,\; AMP\\
BERT~\cite{devlin2019bert}         & Train & 24--36 & BS$\in$\{32,64,128\},\; CK,\; AMP\\
DCGAN~\cite{radford2016dcgan}        & Train & 8--16  & BS$\in$\{32,64,$\cdots$,512\},\; CK,\; AMP\\
PointNet~\cite{qi2017pointnet}     & Train & 12--20 & BS$\in$\{32,64,$\cdots$,512\},\; CK,\; AMP\\
Transformer~\cite{vaswani2017attention}  & Train & 20--38 & BS$\in$\{32,64,$\cdots$,512\},\; CK,\,AMP\\
LSTM~\cite{hochreiter1997lstm}         & Train & 4--12  & BS$\in$\{32,64,$\cdots$,4096\},\; CK,\; AMP\\
NCF-NeuMF~\cite{he2017ncf}    & Train & 4--16  & BS$\in$\{32,64,$\cdots$,8192\},\; CK,\; AMP\\
PPO~\cite{schulman2017ppo}          & Train & 8--24  & BS$\in$\{32,64,$\cdots$,4096\},\; CK,\; AMP\\
\midrule
DeepSeek-1.5b~\cite{deepseek2024llm} & Infer & 16--36  & BS$\in$\{1,2,4\},\; SL$\in$\{4K,8K,16K\},\; PC,\; MU\,0.4--0.9\\
Qwen-1.7b~\cite{yang2024qwen2}     & Infer & 16--36  & BS$\in$\{1,2,4\},\; SL$\in$\{4K,8K,16K\},\; PC,\; MU\,0.4--0.9\\
Mistral-7b~\cite{jiang2023mistral}    & Infer & 28--36 & BS$\in$\{1,2,4\},\; SL$\in$\{1K,2K,4K\},\; PC,\; MU\,0.7--0.9\\
LLaMA-3-7b~\cite{grattafiori2024llama3}   & Infer & 28--36 & BS$\in$\{1,2,4\},\; SL$\in$\{1K,2K,4K\},\; PC,\; MU\,0.7--0.9\\
\bottomrule
\end{tabular}
\par\smallskip\noindent
{\footnotesize BS\,=\,Batch Size;\; SL\,=\,Sequence Length;\; CK\,=\,Activation Checkpointing\,($\pm$);\;
AMP\,=\,Automatic Mixed Precision\,($\pm$);\; PC\,=\,Prefix Caching\,($\pm$);\;
MU\,=\,\texttt{gpu\_memory\_utilization} (vLLM), sampled at 0.05 intervals.\;
$\pm$ denotes an on/off toggle;\; Knob values form a Cartesian product of all listed options.}
\end{table}

\mypara{Baselines.} We compare against two representative schedulers: \textbf{Volcano}~\cite{volcano2024}, a widely-adopted Kubernetes-native batch scheduler that provides gang scheduling, queue management, and fair-share policies but treats job resource requests as fixed and does not perform co-location optimization; and \textbf{Lucid}~\cite{hu2023lucid}, a state-of-the-art DL training scheduler that uses lightweight profiling, indolent resource packing, and priority-based scheduling to minimize average JCT without modifying training code. Volcano represents production-grade static scheduling, while Lucid represents the best available elastic training scheduler.

\mypara{Baseline justification.}
Table~\ref{tab:baseline_comparison} positions \SystemName against recent co-location systems. SIRIUS~\cite{wang2025sirius}, Mudi~\cite{chen2025mudi}, and GPUColo~\cite{chen2024gpucolo} are the closest co-location-aware comparators; however, none have publicly available implementations or standardized benchmark suites, and each targets a different workload mix (SIRIUS: online inference priority; Mudi: SLO-constrained inference batching; GPUColo: latency-guaranteed GPU-local sharing). To provide the fairest comparison possible, we select Volcano and Lucid--whose open-source implementations are mature and reproducible--as anchors representing the two ends of the scheduling spectrum (static vs.\ elastic), and include qualitative feature comparison against co-location systems below. We encourage future work to establish standardized co-location benchmarks.

\begin{table}[pos=t]
\centering\footnotesize
\caption{Qualitative comparison with representative schedulers. ``Config.\ reshape'' denotes whether the scheduler modifies intra-job configuration knobs (batch size, checkpointing, precision); ``Interf.\ pred.'' denotes whether interference is predicted before co-location; ``Cluster-wide'' denotes cluster-level multi-resource optimization (vs.\ GPU-local).}
\label{tab:baseline_comparison}
\begin{tabular}{@{}lcccccc@{}}
\toprule
\textbf{System} & \textbf{Co-location.} & \textbf{Config. reshape} & \textbf{Interf. pred} & \textbf{Cluster-wide} & \textbf{Train+Infer} & \textbf{Open source} \\
\midrule
Volcano~\cite{volcano2024}   & \texttimes & \texttimes & \texttimes & \checkmark & \texttimes & \checkmark \\
Lucid~\cite{hu2023lucid}     & \texttimes & Partial    & \texttimes & \checkmark & \texttimes & \checkmark \\
SIRIUS~\cite{wang2025sirius} & \checkmark & \texttimes & \texttimes & \texttimes & \checkmark & \texttimes \\
Mudi~\cite{chen2025mudi}     & \checkmark & \texttimes & \checkmark & \texttimes & \checkmark & \texttimes \\
GPUColo~\cite{chen2024gpucolo}& \checkmark & \texttimes & \texttimes & \texttimes & \checkmark & \texttimes \\
SMore~\cite{liu2025smore}    & \checkmark & \texttimes & \checkmark & \texttimes & Partial    & \texttimes \\
\midrule
\textbf{\SystemName}         & \checkmark & \checkmark & \checkmark & \checkmark & \checkmark &--\\
\bottomrule
\end{tabular}
\par\smallskip\noindent
{\footnotesize ``Partial'' for Lucid: adjusts GPU count but not intra-job knobs. ``Partial'' for SMore: manages training throughput but not inference throughput.}
\end{table}

\mypara{Metrics.} (1) Average JCT (job completion time, submission to finish); (2) Cluster throughput (total useful work per unit time, normalized); (3) GPU SM utilization; (4) Throughput attainment (fraction of inference tasks meeting throughput targets; fraction of training jobs meeting throughput targets); (5) GPU instance count required for a given workload mix.

\begin{figure}[pos=t]
\centering
\begin{subfigure}[t]{0.48\columnwidth}
  \centering
  \includegraphics[width=\linewidth]{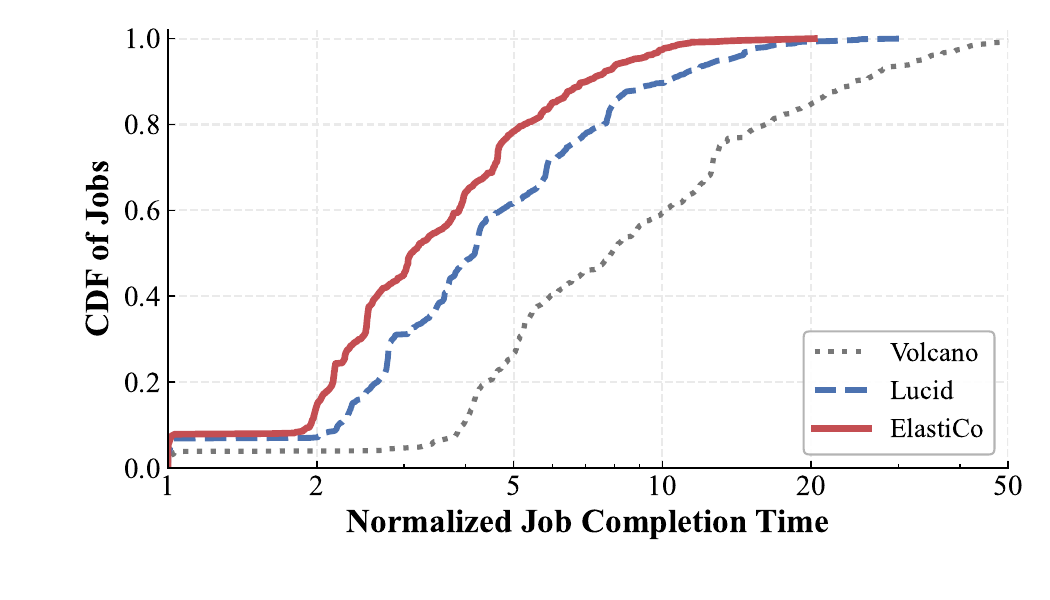}
  \caption{CDF of normalized JCT. \SystemName achieves $2.5\times$ lower median JCT than Volcano.}
  \label{fig:cdf_jct}
\end{subfigure}
\hfill
\begin{subfigure}[t]{0.48\columnwidth}
  \centering
  \includegraphics[width=\linewidth]{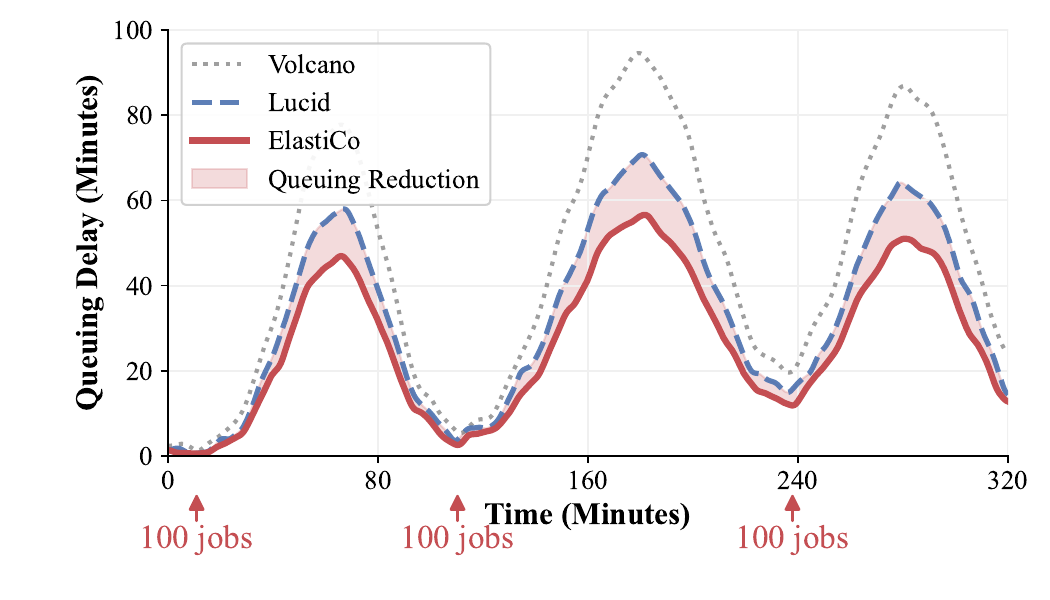}
  \caption{Per-job queuing delay over time. ElastiCo significantly reduces queuing delay and prevents backlog accumulation throughout the workload arrival period.}
  \label{fig:queue_venus}
\end{subfigure}
\caption{Job-level performance comparison on the 64-GPU testbed under mixed workload.}
\label{fig:jct_and_queue}
\end{figure}

\subsection{Overall Performance}
\label{sec:evaluation:comparison}

Figure~\ref{fig:cdf_jct} shows the CDF of normalized JCT for training jobs. \SystemName reduces average JCT by 2.94$\times$ compared to Volcano and by roughly 1.35$\times$ compared to Lucid, while achieving approximately 2.5$\times$ lower median JCT than Volcano. The improvement is driven by two factors: (i) RST enables training jobs to start on fewer GPUs rather than waiting for full allocation, reducing queuing time; (ii) ESP dynamically reconfigures jobs as resources become available, recovering throughput without user intervention.

Figure~\ref{fig:queue_venus} shows per-job queuing delay as a function of arrival order. Volcano exhibits rapidly growing queuing delays as the cluster fills, while Lucid mitigates this partially through elastic packing. ElastiCo significantly reduces queuing delay and mitigates backlog growth even for late-arriving jobs.

\begin{figure}[pos=htbp]
\centering
\includegraphics[width=\columnwidth,height=0.35\textheight,keepaspectratio]{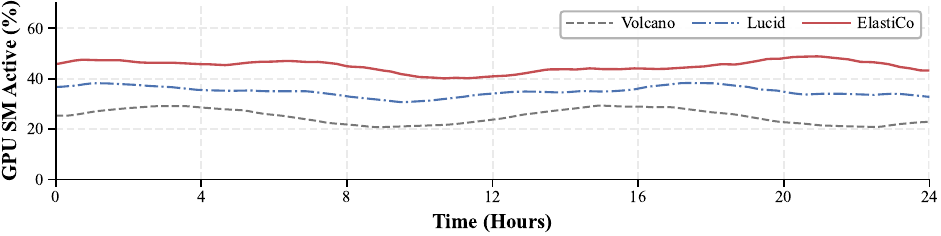}
\caption{Cluster-wide GPU SM Active over a 24-hour trace replay. ElastiCo sustains around 40--50\% utilization, compared to roughly 20--30\% for Volcano.}
\label{fig:gpu_util_time}
\end{figure}

Figure~\ref{fig:gpu_util_time} shows GPU utilization over a 24-hour trace replay. \SystemName raises average SM utilization from 25\% (Volcano) to 46\%, corresponding to a 1.84$\times$ improvement, while Lucid reaches 38\%. The utilization gain is sustained throughout the day: during off-peak hours, training jobs expand to use GPUs vacated by reduced inference demand; during peak hours, RST compresses training configurations to free capacity for inference.

Table~\ref{tab:main_results} summarizes the aggregate comparison across the main metrics. Relative to Volcano, ElastiCo reduces average JCT to 0.34$\times$ and improves cluster throughput to 2.02$\times$, while raising average GPU SM utilization from 25\% to 46\%. Compared to Lucid, ElastiCo further improves both JCT and throughput, indicating that configuration-aware reshaping and interference-aware co-location provide benefits beyond elastic packing alone.

\SystemName also reduces the number of GPU instances required to serve a given mix of training and inference workloads. By co-locating complementary workloads and reshaping configurations to fit available capacity, \SystemName achieves a 44\% reduction in additional GPU instances compared to static partitioning--equivalent to serving the same aggregate workload on substantially fewer devices.

\begin{table}[pos=t]
\centering
\caption{Main comparison results (64-GPU testbed, 24-hour mixed workload). Avg JCT and throughput are normalized to Volcano. Attainment measures the fraction of jobs meeting their throughput target.}
\label{tab:main_results}
\begin{tabular}{@{}lcccc@{}}
\toprule
\textbf{System} & \textbf{Avg JCT} & \textbf{Throughput} & \textbf{SM Util.\,(\%)} & \textbf{Attainment\,(\%)} \\
\midrule
Volcano          & 1.00$\times$ & 1.00$\times$ & 25 & 100\\
Lucid            & 0.46$\times$ & 1.76$\times$ & 38 & 100\\
\midrule
\textbf{\SystemName} & \textbf{0.34$\times$} & \textbf{2.02$\times$} & \textbf{46} & \textbf{98.3}\\
\bottomrule
\end{tabular}
\par\smallskip\noindent
{\footnotesize Volcano and Lucid achieve 100\% attainment trivially by running jobs in isolation (no co-location interference). \SystemName's 98.3\% attainment reflects that IAC successfully prevents harmful co-locations, with the 1.7\% shortfall arising from transient interference during configuration transitions.}
\end{table}

\subsection{Ablation Study}
\label{sec:evaluation:ablation}

To isolate each module's contribution, we disable one component at a time while keeping the rest intact. When RST is disabled, every job runs under a fixed default configuration; when ESP is disabled, a greedy first-fit policy replaces the pricing mechanism; when IAC is disabled, co-location is admitted without interference checking; when PDS is disabled, scheduling decisions are applied without queue-aware capacity reservation or coordinated reconfiguration. Table~\ref{tab:ablation} summarizes the results. 

\begin{table}[pos=t]
\centering
\caption{Ablation study: impact of disabling each \SystemName component.}
\label{tab:ablation}
\begin{tabular}{@{}lcccc@{}}
\toprule
\textbf{Configuration} & \textbf{Avg JCT} & \textbf{Throughput} & \textbf{Mem Util.\%} & \textbf{SM Util.\%} \\
\midrule
\textbf{Full \SystemName} & \textbf{1.00$\times$} & \textbf{1.00$\times$} & \textbf{82.6} & \textbf{46.0} \\
w/o RST          & 1.53$\times$ & 0.77$\times$ & 73.1 & 35.4 \\
w/o ESP          & 1.25$\times$ & 0.82$\times$ & 77.4 & 38.9 \\
w/o IAC          & 1.08$\times$ & 0.92$\times$ & 80.8 & 43.1 \\
w/o PDS          & 1.17$\times$ & 0.89$\times$ & 79.2 & 40.6 \\
\bottomrule
\end{tabular}
\end{table}

\begin{table}[pos=htbp]
\centering
\caption{Representative co-location pairs and observed interference.
Decay denotes the fraction of solo throughput retained under co-location
($1.0$\,=\,no degradation).}
\label{tab:interference_cases}
\begin{tabular}{@{}llccc@{}}
\toprule
\textbf{Severity} & \textbf{Workload Pair (A\,/\,B)} & \textbf{Decay}$_A$ & \textbf{Decay}$_B$ & \textbf{Type} \\
\midrule
\multirow{3}{*}{Severe}
  & Qwen-1.7b$^{\text{PC}}$\,/\,ResNet-50$^{\text{CK}}$      & 0.48 & 0.55 & I\,+\,T \\
  & BERT$^\circ$\,/\,BERT$^{\text{CK}}$                                & 0.47 & 0.57 & T\,+\,T \\
  & DeepSeek-1.5b$^{\text{PC}}$\,/\,ResNet-50$^{\text{CK}}$   & 0.50 & 0.56 & I\,+\,T \\
\midrule
\multirow{4}{*}{High}
  & Qwen-1.7b$^{\text{PC}}$\,/\,ResNet-50$^\circ$                     & 0.52 & 0.57 & I\,+\,T \\
  & PPO$^\circ$\,/\,ResNet-50$^\circ$                                          & 0.54 & 0.72 & T\,+\,T \\
  & Qwen-1.7b$^\circ$\,/\,ResNet-50$^{\text{CK}}$                     & 0.66 & 0.57 & I\,+\,T \\
  & DeepSeek-1.5b$^\circ$\,/\,ResNet-50$^{\text{CK}}$                 & 0.67 & 0.61 & I\,+\,T \\
\midrule
\multirow{6}{*}{Moderate}
  & Qwen-1.7b$^\circ$\,/\,ResNet-50$^\circ$                                   & 0.71 & 0.68 & I\,+\,T \\
  & DeepSeek-1.5b$^\circ$\,/\,PPO$^\circ$                                     & 0.56 & 0.89 & I\,+\,T \\
  & LSTM$^\circ$\,/\,ResNet-50$^{\text{CK}}$                          & 0.70 & 0.70 & T\,+\,T \\
  & LSTM$^\circ$\,/\,NCF$^\circ$                                              & 0.75 & 0.55 & T\,+\,T \\
  & Qwen-1.7b$^{\text{PC}}$\,/\,DeepSeek-1.5b$^{\text{PC}}$   & 0.60 & 0.62 & I\,+\,I \\
  & LLaMA-3-7b$^{\text{PC}}$\,/\,Mistral-7b$^{\text{PC}}$     & 0.58 & 0.61 & I\,+\,I \\
\midrule
\multirow{2}{*}{Low}
  & Qwen-1.7b$^\circ$\,/\,NCF$^\circ$                                         & 0.93 & 0.95 & I\,+\,T \\
  & BERT$^{\text{CK}}$\,/\,DeepSeek-1.5b$^\circ$                      & 0.95 & 0.96 & T\,+\,I \\
\midrule
\multirow{3}{*}{Near-zero}
  & NCF$^\circ$\,/\,PPO$^\circ$                                               & 0.97 & 0.98 & T\,+\,T \\
  & BERT$^\circ$\,/\,PPO$^\circ$                                              & 0.98 & 0.98 & T\,+\,T \\
  & NCF$^\circ$\,/\,BERT$^\circ$                                              & 0.99 & 0.99 & T\,+\,T \\
\bottomrule
\end{tabular}
\par\smallskip\noindent
{\footnotesize 
T\,=\,Training,\; I\,=\,Inference.\;
$^\circ$\,=\,Default configuration;\;
PC\,=\,PrefixCaching;\; CK\,=\,Activation Checkpointing.}
\end{table}

\mypara{RST Unlocks the Co-location Space.}
Disabling RST causes the largest degradation: JCT increases by 53\% and throughput drops by 23\%. The reason is visible in the interference data (Table~\ref{tab:interference_cases}): many workload pairs fall into a moderate-conflict regime (decay $\approx$ 0.6--0.9) that is infeasible under rigid configurations but becomes feasible once RST reshapes resource demands. For example, enabling activation checkpointing for a BERT job reduces its memory footprint from 36\,GB to 24\,GB, opening sufficient headroom for co-location with an inference task. GPU memory utilization drops from 82.6\% to 73.1\% without RST, confirming that fixed configurations leave substantial capacity stranded. \textbf{Insight:} The dominant source of improvement is demand reshaping, not job reordering.

\mypara{IAC Prevents Catastrophic Interference.}
Removing IAC only slightly reduces throughput (about 8\%) but leads to noticeably worse co-location quality, because high-conflict pairs are admitted without restriction and can cause severe slowdown for individual jobs.  \textbf{Insight:} IAC is essential for avoiding localized performance collapse rather than improving average throughput.

\mypara{ESP Improves Global Allocation Efficiency.}
Replacing ESP with a greedy policy reduces throughput by 18\%. Interference is highly heterogeneous (Table~\ref{tab:interference_cases}), with some pairs nearly interference-free and others significantly degraded. Greedy decisions fail to account for this variability, while ESP prioritizes globally efficient, low-interference combinations. \textbf{Insight:} Exploiting heterogeneity requires global coordination.

\mypara{PDS as the Orchestration Layer.}
PDS integrates configuration selection, interference feedback, and runtime adaptation into a unified loop. Disabling it leads to moderate degradation, as decisions become unstable under dynamic workloads. Without coordinated handling of reconfiguration and interference feedback, the system suffers from inconsistent placement and unnecessary adjustments. \textbf{Insight:} Effective scheduling requires not only good decisions, but consistent execution.

\subsection{Case Study}
\label{sec:evaluation:case}

\begin{figure}[pos=t]
\centering
\begin{subfigure}[t]{0.40\columnwidth}
  \centering
  \includegraphics[width=\linewidth]{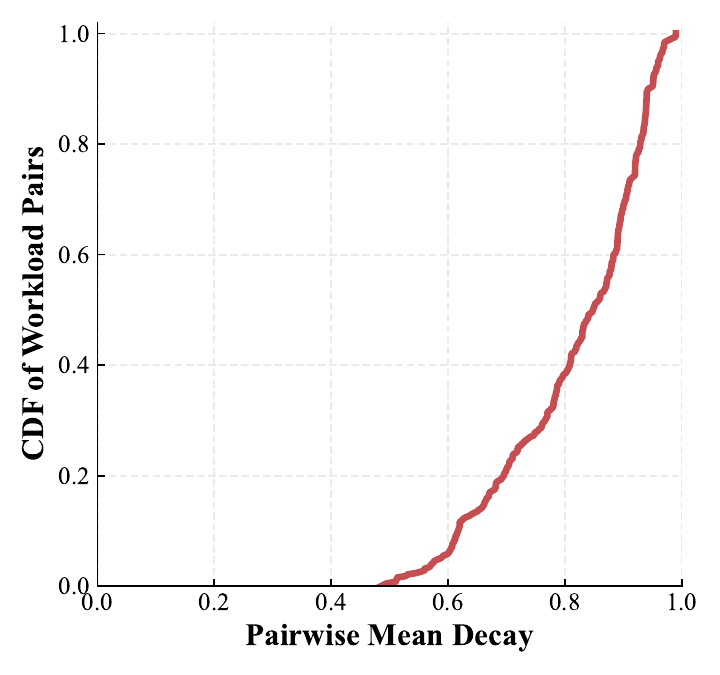}
  \caption{CDF of pairwise mean performance decay across workload pairs.}
  \label{fig:interference_cdf}
\end{subfigure}
\hfill
\begin{subfigure}[t]{0.56\columnwidth}
  \centering
  \includegraphics[width=\linewidth]{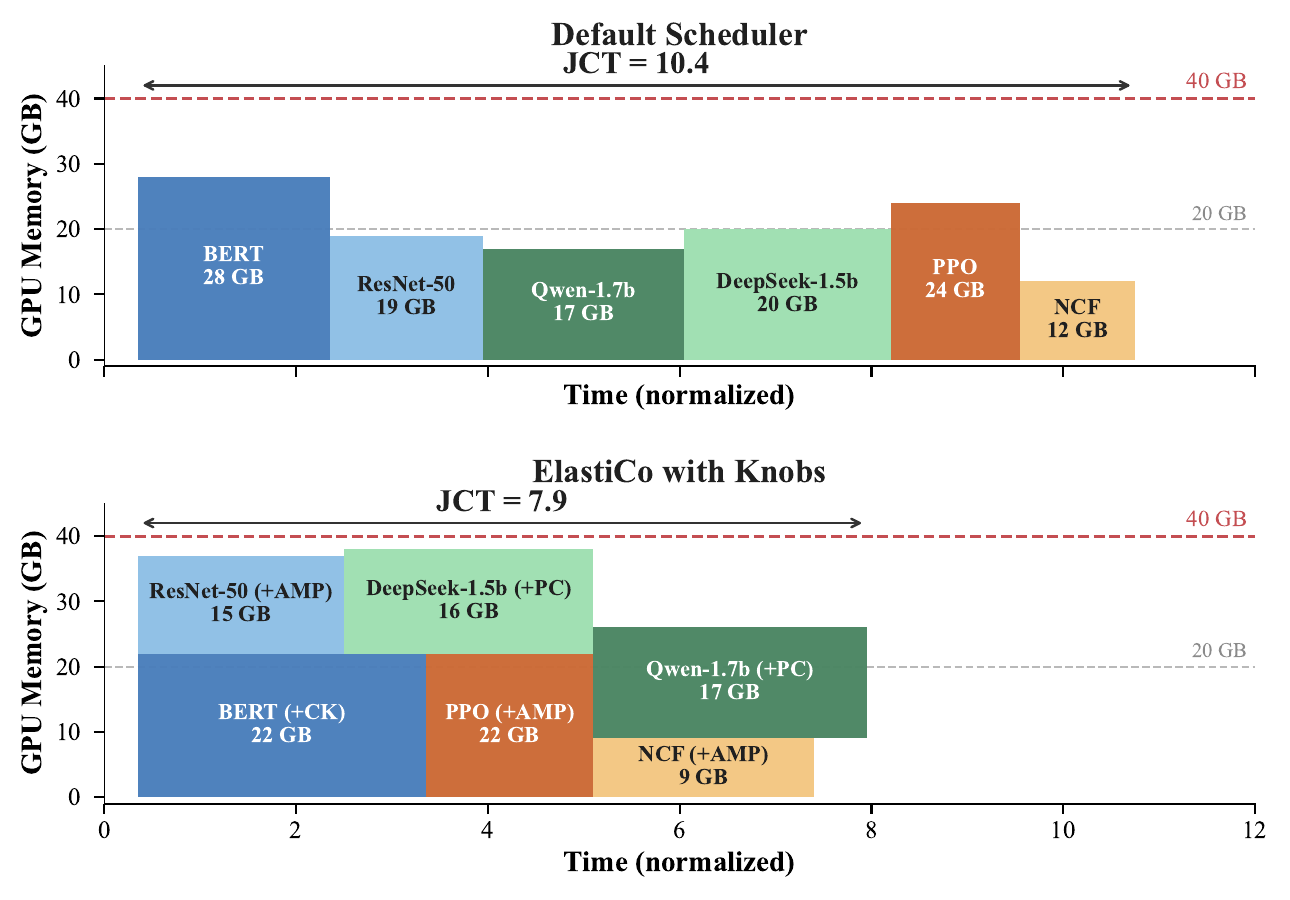}
  \caption{Resource demand timelines under co-location for representative workload pairs.}
  \label{fig:interference_profile}
\end{subfigure}
\caption{Interference characteristics of workload co-location.}
\label{fig:interference_structure}
\end{figure}

To better understand where the performance gains of \SystemName come from, we combine the global interference structure with concrete co-location examples. 
Figure~\ref{fig:interference_structure}(a) characterizes the overall distribution of pairwise interference, while Figure~\ref{fig:interference_structure}(b) presents representative execution timelines for selected workload pairs.

\mypara{Interference structure.}

\begin{figure}[pos=htbp]
\centering
\includegraphics[width=\columnwidth,keepaspectratio]{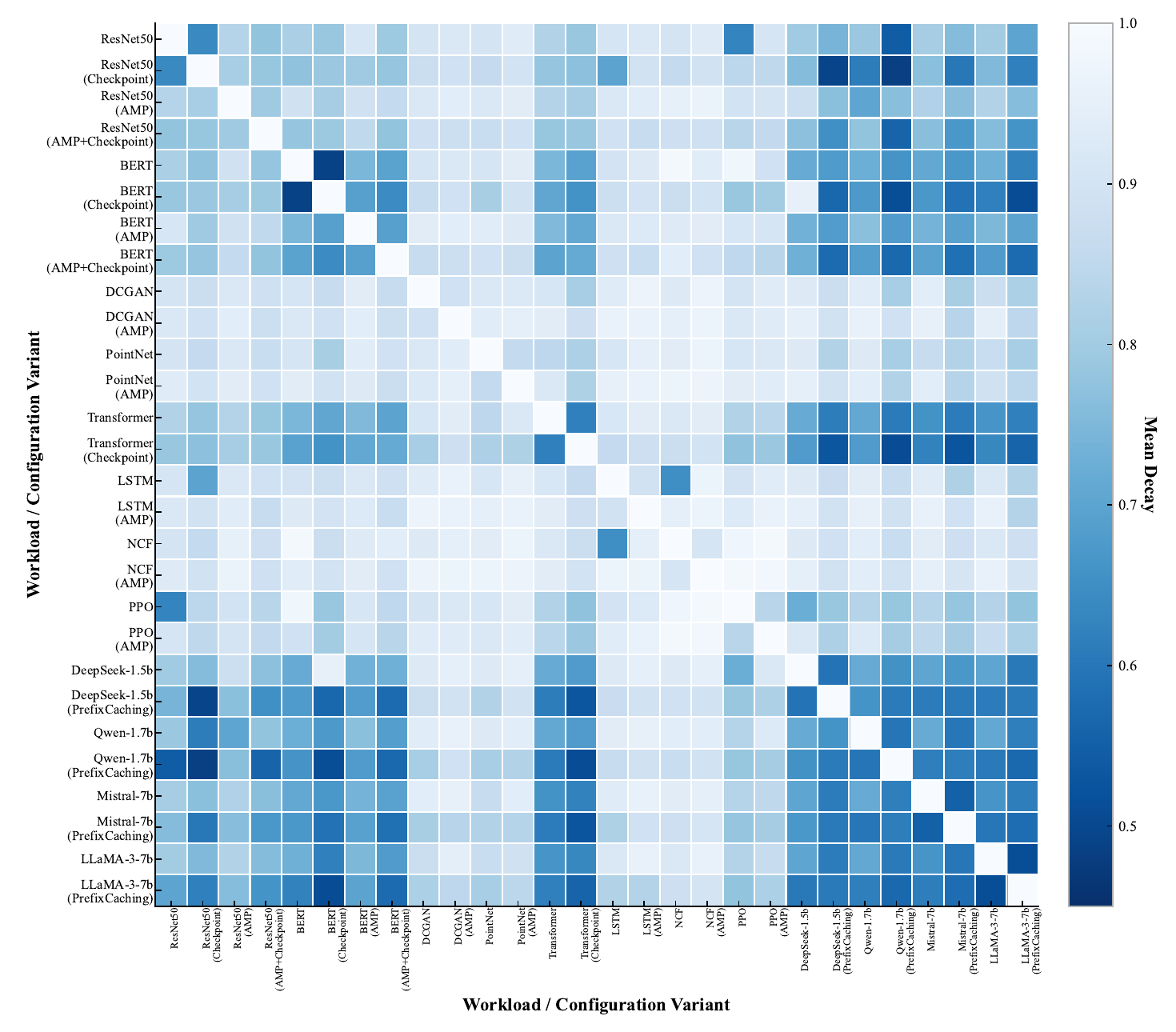}
\caption{Heatmap of pairwise co-location interference across workload variants.}
\label{fig:hitmap}
\end{figure}

As shown in Figure~\ref{fig:hitmap}, co-location interference exhibits strong heterogeneity: the observed performance decay ranges from near-zero impact to significant slowdown.
Moreover, interference is not solely determined by workload type, but depends critically on their configuration-dependent resource demands, including memory footprint and compute intensity.
This suggests that co-location feasibility and efficiency are fundamentally shaped by how resource demands align under specific configurations.

Motivated by this observation, we analyze three representative scenarios that illustrate distinct mechanisms through which \SystemName improves scheduling quality.

\mypara{Case 1: Infeasible co-location resolved by RST reshaping.}
Under the default scheduler, BERT (36\,GB) and ResNet-50 (24\,GB) each run with their peak-throughput configuration, totaling 60\,GB--well beyond the 40\,GB capacity of a single A100. The scheduler serializes them, leaving one GPU idle while the other runs.
\SystemName enables CK for BERT (reducing its footprint to 24\,GB) and AMP for ResNet-50 (reducing to 14\,GB). The combined 38\,GB fits within device capacity, enabling concurrent execution and reducing the pair's makespan by 41\%.
\textbf{Insight:} rigid configurations create hard infeasibility; RST converts infeasible pairs into feasible co-locations by reshaping resource demands.

\mypara{Case 2: Avoidable queuing resolved by ESP pricing.}
Consider a burst of four training jobs (ResNet-50, BERT, PPO, NCF) arriving simultaneously on a two-GPU node. Under a greedy first-fit policy, the scheduler assigns the two largest jobs (BERT, ResNet-50) to separate GPUs and queues PPO and NCF. ESP, by contrast, raises GPU memory prices in response to the overload, steering all four jobs toward lower-memory configurations (smaller batch sizes, AMP enabled). This allows all four jobs to start immediately via two co-located pairs, eliminating the queuing delay entirely. The 30-minute queuing time observed under the greedy policy is reduced to zero, at a cost of 8--12\% per-job throughput reduction from the configuration downgrade--a trade-off that reduces average JCT by 35\%.
\textbf{Insight:} ESP's price signals dynamically steer jobs toward configurations that collectively fit, converting queuing delays into controlled throughput trade-offs.

\mypara{Case 3: Configuration-enabled dense packing.}
Figure~\ref{fig:interference_structure}(b) compares the execution timelines of six representative workloads under the default scheduler and \SystemName.
Under the baseline scheduler, all jobs run with fixed configurations and rigid resource allocations.
Due to their high GPU memory footprints, the combined demand of multiple jobs exceeds device capacity, making concurrent execution infeasible.
As a result, the scheduler is forced to execute jobs in a largely serialized manner, with minimal overlap dictated by residual capacity rather than scheduling decisions.
This leads to significant idle periods and an overall job completion time (JCT) of 10.4.
\SystemName overcomes this limitation by actively reshaping workload resource demands through configuration knobs.
Specifically, checkpointing (CK) is enabled for BERT, AMP is applied to ResNet-50, PPO, and NCF, and prefix caching (PC) is enabled for the inference workloads (Qwen-1.7b and DeepSeek-1.5b).
These transformations reduce per-job memory footprints, allowing the aggregate demand to fit within device capacity. Multiple jobs become jointly schedulable, enabling substantial temporal overlap. The overall JCT is reduced from 10.4 to 7.9--a 24\% improvement from configuration-level resource reshaping alone.
\textbf{Insight:} the combination of RST reshaping, ESP pricing, and IAC interference checking produces a densely packed schedule that no single mechanism can achieve in isolation.

\subsection{Interference Prediction Accuracy}
\label{sec:evaluation:prediction}

The IAC module's utility depends on the accuracy of its slowdown predictions. We evaluate the DNN-based predictor using 5-fold cross-validation on the co-location dataset collected from the testbed, which comprises pairwise measurements across all workload combinations listed in Table~\ref{tab:workloads} under multiple configuration variants.

\begin{table}[pos=t]
\centering
\caption{IAC prediction accuracy (5-fold cross-validation). MAPE\,=\,mean absolute percentage error; $R^2$\,=\,coefficient of determination.}
\label{tab:prediction_accuracy}
\begin{tabular}{@{}lccc@{}}
\toprule
\textbf{Co-location Type} & \textbf{MAPE (\%)} & \textbf{$R^2$} & \textbf{Samples} \\
\midrule
Training + Training   & 6.8  & 0.91 & 384 \\
Training + Inference  & 8.2  & 0.88 & 512 \\
Inference + Inference & 7.5  & 0.89 & 192 \\
\midrule
\textbf{Overall}      & \textbf{7.6} & \textbf{0.89} & \textbf{1088} \\
\bottomrule
\end{tabular}
\end{table}

Table~\ref{tab:prediction_accuracy} reports per-category and overall accuracy. The predictor achieves a mean absolute percentage error (MAPE) of 7.6\% across all co-location types, with an $R^2$ of 0.89. Training--training pairs are predicted most accurately (6.8\% MAPE), likely because training workloads exhibit more stable, periodic resource consumption. Training--inference pairs show slightly higher error (8.2\%), reflecting the greater variability introduced by continuous-batching dynamics in inference workloads. Importantly, prediction errors are small relative to the interference magnitudes observed in practice (Table~\ref{tab:interference_cases}): the difference between a safe pair (decay $>0.9$) and a harmful pair (decay $<0.6$) spans 30--40 percentage points, while the predictor's typical error is under 8 percentage points. This margin is sufficient for the scheduler to reliably distinguish safe from harmful co-locations.

We additionally compare the DNN predictor against two alternatives: a gradient-boosted decision tree (XGBoost) and a linear regression baseline. The DNN achieves 7.6\% MAPE, compared to 9.4\% for XGBoost and 18.7\% for linear regression. The DNN's advantage is most pronounced for moderate-interference pairs (decay 0.6--0.9), where smooth interpolation in the feature space is critical for scheduling-relevant distinctions--consistent with the design rationale in \S\ref{sec:methodology:iac}.

\subsection{Scalability}
\label{sec:evaluation:scalability}

To assess \SystemName's behavior beyond the 64-GPU testbed, we use the discrete-event simulator described in \S\ref{sec:evaluation:setup} to evaluate performance at 64, 128, 256, and 512 GPUs. The simulator is calibrated against testbed measurements and validated to within 7\% of physical-cluster results for JCT and throughput metrics. Workload arrival rates are scaled proportionally to cluster size to maintain a consistent load factor.

\begin{table}[pos=t]
\centering
\caption{Scalability: \SystemName performance at increasing cluster scale (simulation). All metrics normalized to the Volcano baseline at the same scale.}
\label{tab:scalability}
\begin{tabular}{@{}rcccc@{}}
\toprule
\textbf{GPUs} & \textbf{Avg JCT} & \textbf{Throughput} & \textbf{SM Util.\,(\%)} & \textbf{Sched.\,Latency\,(ms)} \\
\midrule
64   & 0.34$\times$ & 2.02$\times$ & 46 & 4.2 \\
128  & 0.36$\times$ & 1.95$\times$ & 44 & 6.8 \\
256  & 0.38$\times$ & 1.89$\times$ & 43 & 9.1 \\
512  & 0.41$\times$ & 1.82$\times$ & 41 & 13.7 \\
\bottomrule
\end{tabular}
\end{table}

Table~\ref{tab:scalability} shows that \SystemName maintains substantial improvements over the baseline across all scales. At 512 GPUs, \SystemName still reduces average JCT by 2.44$\times$ and improves throughput by 1.82$\times$ relative to Volcano. The modest degradation at larger scales (from 2.02$\times$ at 64 GPUs to 1.82$\times$ at 512 GPUs) arises because the ESP t\^{a}tonnement converges more slowly as the number of interacting jobs grows; however, the per-epoch scheduling latency remains well under the 5-second epoch interval even at 512 GPUs (13.7\,ms), confirming that the $O(M \cdot I \cdot N \cdot \bar{K})$ complexity analysis in \S\ref{sec:methodology:pds} translates to practical scalability.

We also examine sensitivity to load intensity by varying the job arrival rate at the 128-GPU scale. At low load (50\% of peak arrival rate), all schedulers perform comparably because resources are abundant. As load increases to 100\% and 150\% of peak, \SystemName's advantage widens: the JCT improvement over Volcano grows from 1.8$\times$ at 50\% load to 2.9$\times$ at 150\% load. Under heavy contention, the combination of RST-enabled reshaping and ESP-driven pricing becomes especially valuable, as it enables the scheduler to pack more jobs onto the same hardware by finding interference-safe configuration combinations that rigid schedulers cannot access.

\subsection{System Overhead}
\label{sec:evaluation:overhead}

We quantify four sources of overhead introduced by \SystemName.

\mypara{Profiling overhead.}
Initial profile generation takes 3--5 minutes per (model architecture, configuration) pair, including warm-up and measurement. However, due to the profile cache (\S\ref{subsec:profiling}), the amortized profiling overhead is low: 94\% of job arrivals in our evaluation match an existing profile, requiring zero additional profiling time. For the remaining 6\%, profiling runs concurrently on a dedicated profiling GPU and does not delay other scheduling decisions.

\mypara{Scheduling latency.}
Table~\ref{tab:overhead_scheduling} breaks down the per-epoch scheduling latency by component. The end-to-end latency is dominated by the ESP t\^{a}tonnement (typically 3--5 iterations) and IAC prediction; bin-packing and PDS coordination add minimal overhead. At 64 GPUs with 30 active jobs, the total scheduling latency is 4.2\,ms; at 512 GPUs with 200 active jobs, it grows to 13.7\,ms--both negligible relative to the 5-second scheduling epoch.

\begin{table}[pos=t]
\centering
\caption{Per-epoch scheduling latency breakdown (64-GPU testbed, 30 active jobs).}
\label{tab:overhead_scheduling}
\begin{tabular}{@{}lr@{}}
\toprule
\textbf{Component} & \textbf{Latency (ms)} \\
\midrule
ESP t\^{a}tonnement (3--5 iterations) & 2.1 \\
IAC prediction (all pairs)             & 1.2 \\
Bin-packing                            & 0.5 \\
PDS coordination                       & 0.4 \\
\midrule
\textbf{Total}                         & \textbf{4.2} \\
\bottomrule
\end{tabular}
\end{table}

\mypara{Reconfiguration overhead.}
Configuration transitions vary in scope and cost. Flag toggles (e.g., enabling prefix caching) take effect immediately via a side-channel RPC to the data-plane agent (sub-millisecond). Engine restarts (e.g., changing continuous-batching parameters in vLLM) require draining in-flight requests and relaunching the serving process (5--15\,s). Checkpoint--resume transitions (e.g., switching activation recomputation or precision mode) serialize model state to persistent storage, terminate the pod, and relaunch under the new configuration (15--45\,s depending on model size). In the 24-hour evaluation trace, the scheduler triggered an average of 2.3 checkpoint--resume transitions per hour per GPU, with each transition consuming less than 0.3\% of total GPU-hours.

\mypara{Interference monitoring overhead.}
The TDOP data-plane agent consumes less than 1\% additional CPU and negligible GPU overhead, as it relies on passive DCGM sampling (1-second intervals) and lightweight \texttt{LD\_PRELOAD} wrappers that add no measurable latency to CUDA kernel launches. The memory footprint of the per-process ring buffer is 4\,MB.

\section{Related Work}
\label{sec:related_work}

\mypara{Training-inference co-location.}
Co-locating training and inference workloads on a shared GPU cluster is an emerging strategy for improving cluster utilization.
SIRIUS~\cite{wang2025sirius} prioritizes inference tasks with unrestricted GPU access and runs training on leftover resources, using millisecond-level gradient-aware memory adjustment and SLO-aware reallocation for fast GPU memory handover, but treats each job's resource configuration as fixed, which limits packing flexibility when cluster-wide resource pressure changes.
SMore~\cite{liu2025smore} enhances GPU utilization through serverless-based co-location scheduling with a degradation prediction model, yet focuses exclusively on training-side performance and does not manage offline inference throughput.
ConServe~\cite{qiao2025conserve} co-serves latency-critical online requests with latency-tolerant offline tasks on shared GPUs through fine-grained harvesting--token-level scheduling, layer-wise preemption, and incremental KV cache management--but targets a single LLM serving engine rather than cluster-wide training-inference co-location.
GPUColo~\cite{chen2024gpucolo} provides latency-guaranteed co-location through a two-tier mechanism--an outer tier that dynamically adjusts MPS thread percentages via spatial sharing and an inner tier that periodically sleeps training processes for prompt inference latency control--but confines adaptation to GPU-local knobs without cluster-wide multi-resource optimization across both workload classes.
Mudi~\cite{chen2025mudi} multiplexes inference services with training tasks through spatial sharing, using piece-wise linear profiling to quantify resource interference and adaptive batching to handle dynamic workloads, but its co-location policy optimizes GPU-percentage allocation without reshaping per-job configurations. Llumnix~\cite{sun2024llumnix} reschedules requests across LLM serving instances at runtime via live migration to improve load balancing, reduce fragmentation, and differentiate priorities, but operates entirely within the inference serving layer without addressing training--inference co-location.
LeMix~\cite{li2025lemix} co-locates LLM serving and training on shared multi-GPU nodes, using offline profiling and runtime scheduling to adapt resource allocation based on workload characteristics and co-execution interference; however, it focuses on single-model retraining scenarios and does not address cluster-wide heterogeneous job scheduling.
In contrast, \SystemName jointly optimizes both training and inference configurations through RST, coordinates resource allocation via market-based pricing (ESP), and feeds interference predictions (IAC) back into the allocation loop--capabilities that none of the above systems integrate simultaneously.

\mypara{Elastic DL cluster scheduling.}
Several schedulers allow dynamic adaptation of job resource allocations.
Pollux~\cite{qiao2021pollux} co-adaptively optimizes both per-job training hyperparameters (batch size, learning rate) and cluster-wide GPU allocation using a goodput metric that combines system throughput with statistical efficiency; however, it targets training jobs exclusively and does not handle inference workloads or co-location interference.
Sia~\cite{jayaram2023sia} introduces a scalable scheduling formulation that matches elastic resource-adaptive jobs to heterogeneous GPU types and counts, with low-overhead throughput-model bootstrapping and the first support for elastic scaling of hybrid parallel (data + pipeline) jobs, but focuses exclusively on training and does not model co-location interference.
Shockwave~\cite{zheng2023shockwave} extends classic market theory to dynamic settings and uses stochastic dynamic programming with future planning to co-optimize fairness and efficiency under dynamic job adaptation; while it shares the market-theoretic spirit with \SystemName's ESP module, it does not support intra-GPU co-location or interference-aware placement.
Lucid~\cite{hu2023lucid} uses lightweight profiling and an indolent packing algorithm to schedule DL training jobs non-intrusively, minimizing average JCT without modifying user code; it targets training workloads exclusively and does not model co-location interference.
PowerFlow~\cite{gu2023powerflow} dynamically allocates GPUs and adjusts power configurations to minimize average JCT under an energy budget, demonstrating energy-efficiency--performance co-optimization in GPU clusters.
Mist~\cite{zhu2025mist} comprehensively co-optimizes memory footprint reduction techniques (checkpointing, offloading, redundancy elimination) alongside parallelism strategies for LLM training via symbolic performance analysis and imbalance-aware hierarchical tuning, but targets training-only optimization without co-location considerations.
Earlier systems such as Gandiva~\cite{xiao2018gandiva} (time-slicing and migration) and Tiresias~\cite{gu2019tiresias} (information-agnostic priority scheduling that minimizes average JCT for jobs with unpredictable durations) treat job resource requirements as fixed.
\SystemName extends the elastic scheduling paradigm by formalizing configuration flexibility across both training and inference workloads (via RST) and coupling it with a decentralized pricing mechanism that scales to large clusters.

\mypara{Market-based resource allocation.}
\SystemName's ESP module builds on a rich theoretical foundation in market-based resource allocation.
Kelly et al.~\cite{kelly1998rate} establish proportional fairness and shadow pricing for communication networks, showing that distributed agents responding to price signals can converge to a socially optimal allocation.
Dominant Resource Fairness (DRF)~\cite{ghodsi2011dominant} extends max-min fairness to multiple resource types, providing a fairness baseline for multi-resource environments.
Carbyne~\cite{grandl2016carbyne} applies altruistic resource sharing in multi-tenant clusters.
In the GPU scheduling domain, Shockwave~\cite{zheng2023shockwave} extends the Fisher market framework to dynamic settings and solves the allocation via stochastic dynamic programming with future planning, rather than iterative price adjustment.
\SystemName differs in two key aspects: (i) it employs t\^{a}tonnement-style iterative price updates (Eq.~\ref{eq:price_update}) that naturally accommodate the discrete configuration sets produced by RST, and (ii) it embeds interference penalties from IAC directly into the pricing loop, enabling the market mechanism to account for co-location externalities that conventional market formulations ignore.

\mypara{GPU sharing and interference modeling.}
Intra-GPU sharing mechanisms determine \textit{how} multiple workloads can safely share a single GPU.
AntMan~\cite{xiao2020antman} co-designs cluster scheduling with DL frameworks to dynamically scale GPU memory and computation, enabling opportunistic co-execution of multiple jobs on shared GPUs without interference.
Salus~\cite{yu2020salus} enables fine-grained GPU sharing via two primitives--fast job switching and memory sharing--through iteration-level scheduling, and PipeSwitch~\cite{bai2020pipeswitch} pipelines model transmission and GPU execution by exploiting the layered structure of neural networks, achieving millisecond-scale task switching for DL application time-sharing.
NVIDIA MPS~\cite{nvidia2019mps} and MIG~\cite{nvidia2020mig} provide hardware-level spatial sharing primitives.
Orion~\cite{strati2024orion} transparently intercepts GPU kernel launches and schedules work at individual operator granularity, accounting for each operator's compute and memory requirements to minimize interference among co-located workloads.
More recently, ISACPP~\cite{liu2025isacpp} builds an edge-fusion gated graph attention network incorporating DL model structures and GPU types to predict co-location performance, and proposes a multi-stage interference quantification model to identify minimum-interference GPU placements.
USHER~\cite{shubha2024usher} maximizes GPU utilization for ML inference by profiling per-kernel resource requirements, scheduling co-located models with an interference-aware heuristic, and merging operator graphs to reduce cache contention.
KACE~\cite{han2024kace} predicts co-location interference from exclusive kernel-level GPU metrics with minimal profiling overhead, enabling lightweight co-location decisions under CUDA-MPS spatial sharing.
Hu et al.~\cite{hu2024characterization} present an in-depth characterization of LLM development workloads in GPU datacenters, revealing resource utilization imbalances and the impact of frequent job failures at scale.
\SystemName's IAC module uses a DNN-based regressor over hardware-counter and task-level features, which generalizes across heterogeneous model architectures without requiring computation graph information; furthermore, IAC feeds its predictions back into ESP's pricing loop rather than acting as a standalone post-hoc filter.

\mypara{LLM inference and serving.}
A separate line of work optimizes the serving stack for large language models.
vLLM~\cite{kwon2023vllm} introduces PagedAttention for efficient KV-cache memory management.
DistServe~\cite{zhong2024distserve} assigns prefill and decoding to separate GPUs, co-optimizing per-phase resource allocation and parallelism strategies to meet both TTFT and TPOT latency targets.
Splitwise~\cite{patel2024splitwise} splits the compute-intensive prompt computation and memory-intensive token generation phases onto separate machines, enabling phase-specific hardware matching for higher throughput at lower cost.
AlpaServe~\cite{li2023alpaserve} exploits model parallelism not only for scaling large models but also for statistical multiplexing across devices, determining efficient placement and parallelization strategies to reduce serving latency under bursty workloads.
SpotServe~\cite{miao2024spotserve} dynamically adapts LLM parallelization configurations on preemptible cloud instances, using optimal migration planning and stateful inference recovery to maintain serving quality despite frequent instance preemptions.
Vidur~\cite{agrawal2024vidur} provides a high-fidelity simulation framework for LLM inference that models operator performance via profiling and predictive modeling, enabling rapid exploration of parallelization, batching, and scheduling configurations at scale without expensive real-cluster experiments.
DeepServe~\cite{hu2025deepserve} is a serverless AI platform that efficiently serves LLMs at scale in cloud environments, integrating PD-disaggregated and PD-colocated scheduling, NPU-centric execution, and rapid scaling optimizations for production deployment on large Ascend clusters.
SGLang~\cite{zheng2024sglang} introduces RadixAttention for KV cache reuse and compressed finite state machines for structured output decoding, achieving up to $6.4\times{}$ higher throughput on complex LLM programs.
These systems optimize single-workload inference serving within an individual serving engine and are \emph{complementary} to \SystemName, which operates at the cluster scheduling layer to determine resource allocation and co-location decisions above the serving stack.

\section{Conclusion}
\label{sec:conclusion}

This paper presented \SystemName, an elastic co-location framework for multi-tenant GPU clusters that concurrently serve deep learning training and offline LLM inference workloads. Our central thesis is that cluster schedulers must jointly reason about \emph{intra-job configuration flexibility} and \emph{inter-job interference} to unlock the utilization headroom left by static-allocation policies--and that solving either in isolation yields only marginal gains, because the unaddressed dimension becomes the binding constraint.

Guided by this, we introduced three integrated mechanisms: (1)~\emph{Resource Shape Transformation (RST)}, which exposes each job as a family of feasible resource--performance profiles by systematically exploring activation recomputation, micro-batch sizing, mixed-precision modes, and KV-cache configurations--making intra-job configuration flexibility a first-class scheduling dimension for the first time in the training--inference co-location setting; (2)~\emph{Elastic Shadow Pricing (ESP)}, which decomposes the resulting combinatorial multi-resource allocation problem into per-job subproblems via Lagrangian relaxation with Pigouvian interference penalties, scaling to hundreds of concurrent jobs in under 15\,ms; and (3)~\emph{Interference-Aware Co-location (IAC)}, which predicts pairwise slowdown under NVIDIA MPS co-execution using a DNN-based model on hardware-counter and task-level features (7.6\% MAPE) and enforces adaptive tolerance thresholds to bound throughput degradation. The \emph{Phase-Aware Disaggregated Scheduling (PDS)} module orchestrates these components in a closed-loop control cycle with queue-aware capacity reservation and runtime adaptation.

Implemented as Kubernetes-native middleware requiring no modifications to user code, \SystemName reduces average job completion time by up to 2.94$\times$, increases cluster throughput by 2.02$\times$, raises GPU utilization from approximately 25\% to 46\%, and reduces the additional GPU instances required for concurrent offline inference by 44\% compared to static partitioning--all while meeting throughput targets for 98.3\% of jobs. Ablation experiments confirm that each component contributes complementary gains: RST provides the largest gain in scheduling flexibility (53\% JCT increase when disabled), ESP improves global allocation efficiency through heterogeneity-aware coordination, and IAC is essential for preventing localized performance collapse under aggressive GPU sharing.

\mypara{Limitations.}
We identify four limitations that scope the current contribution.
\textit{(1)~Single-GPU configuration scope:} the current RST implementation and evaluation focus on per-GPU configuration knobs (batch size, checkpointing, precision, KV-cache parameters). Although the RST abstraction naturally extends to multi-GPU parallelism strategies (data, tensor, pipeline parallelism), this extension introduces inter-GPU communication modeling and is left for future work.
\textit{(2)~Hardware homogeneity:} all experiments use NVIDIA A100-40GB GPUs; profile accuracy on other GPU generations (e.g., H100, L40S) requires re-profiling.
\textit{(3)~Moderate cluster scale:} physical validation is limited to 64 GPUs, with simulation extending to 512 GPUs. Larger-scale deployments may reveal additional scheduling dynamics.
\textit{(4)~IAC generalization:} the interference predictor is trained on the 12 workloads in our evaluation mix; accuracy may degrade on model architectures with substantially different hardware utilization profiles (e.g., mixture-of-experts or diffusion models).

\mypara{Future work.}
Promising extensions include: (i)~integrating multi-GPU parallelism strategies into RST, enabling the scheduler to jointly select parallelism configurations and per-GPU knobs--this would subsume systems like Mist~\cite{zhu2025mist} into the RST framework; (ii)~supporting heterogeneous multi-generation GPU fleets, where RST profiles must capture hardware-specific performance characteristics; (iii)~incorporating network bandwidth and storage I/O into the resource-shape representation for data-intensive distributed training; and (iv)~adopting online learning for the IAC predictor to handle novel model architectures without offline profiling, enabling continuous adaptation as the workload mix evolves.

\section*{Funding}
This work is supported in part by National Key
R\&D Program of China (Grant No. 2024YFB4505604), in part by the National Natural Science Foundation of China (Grant No. 62402024), in part by the Beijing Natural Science Foundation (Grant No. L241050), in part by the Fundamental Research Funds for the Central Universities, and, last but not least, by Kuaishou Research Fund.

\section*{Declaration of generative AI and AI-assisted technologies in the manuscript preparation process}
During the preparation of this work the author(s) used ChatGPT in order to improve the linguistic clarity and readability of the manuscript. After using this tool/service, the author(s) reviewed and edited the content as needed and take(s) full responsibility for the content of the published article.

\bibliographystyle{elsarticle-num-names}
\bibliography{refs}

\end{document}